\documentclass{sig-alternate} 
\usepackage{mathptmx} 
\usepackage{subcaption}

\newcommand{\ignore}[1]{}
\usepackage{fancyhdr}
\usepackage[normalem]{ulem}
\usepackage[hyphens]{url}
\usepackage{microtype}
\usepackage{braket}
\usepackage{xfrac}
\usepackage{algorithm}
\usepackage{algpseudocode}
\usepackage{algorithmicx}
\usepackage{enumerate}
\usepackage{multirow}
\usepackage[usenames, dvipsnames]{color}

\usepackage{hyperref}

\DeclareMathOperator*{\argmin}{arg\,min}
\renewcommand{\algorithmicrequire}{\textbf{Input:}}
\renewcommand{\algorithmicensure}{\textbf{Output:}}

\newcommand{\iscasubmissionnumber}{496}

\fancypagestyle{firstpage}{
  \fancyhf{}

  \fancyhead[C]{\normalsize{Microsys Journal Submission
      \textbf{\#\iscasubmissionnumber} \\ Confidential Draft: DO NOT DISTRIBUTE}} 
  \fancyfoot[C]{\thepage}
}

\title{\Large{\bf Resource Optimized Quantum Architectures for \\Surface Code Implementations of Magic-State Distillation} \vspace{-0.5in}}

\author{\normalfont{Adam Holmes,\thanks{Corresponding authors: \{adholmes, yongshan\}@uchicago.edu. These two authors contributed equally. }$^{\ \ ,}$\IEEEauthorrefmark{2} Yongshan Ding,\footnotemark[1] $^{\ ,}$\IEEEauthorrefmark{2} Ali Javadi-Abhari,\IEEEauthorrefmark{3} Diana Franklin,\IEEEauthorrefmark{2} Margaret Martonosi,\IEEEauthorrefmark{3} and Frederic T. Chong\IEEEauthorrefmark{2}}\\\IEEEauthorrefmark{2}\it{Department of Computer Science, University of Chicago, Chicago IL 60637, USA}\\\IEEEauthorrefmark{3}\it{Department of Computer Science, Princeton University, Princeton NJ 08540, USA}}

\begin{document}
\maketitle

\begin{abstract}

Quantum computers capable of solving classically intractable problems are under construction, and intermediate-scale devices are approaching completion. Current efforts to design large-scale devices require allocating immense resources to error correction, with the majority dedicated to the production of high-fidelity ancillary states known as magic-states. Leading techniques focus on dedicating a large, contiguous region of the processor as a single ``magic-state distillation factory'' responsible for meeting the magic-state demands of applications. 


In this work we design and analyze a set of optimized factory architectural layouts that divide a single factory into spatially distributed factories located throughout the processor. We find that distributed factory architectures minimize the space-time volume overhead imposed by distillation. Additionally, we find that the number of distributed components in each optimal configuration is sensitive to application characteristics and underlying physical device error rates. More specifically, we find that the rate at which T-gates are demanded by an application has a significant impact on the optimal distillation architecture. We develop an optimization procedure that discovers the optimal number of factory distillation rounds and number of output magic states per factory, as well as an overall system architecture that interacts with the factories. This yields between a 10x and 20x resource reduction compared to commonly accepted single factory designs. Performance is analyzed across representative application classes such as quantum simulation and quantum chemistry.

\end{abstract}


\section{Introduction}\label{sec:Introduction}

 Quantum computers promise to provide computational power required to solve classically intractable problems and have significant impacts in materials science, quantum chemistry, cryptography, communication, and many other fields. Recently, much focus has been placed on constructing and optimizing Noisy Intermediate-Scale Quantum (NISQ) computers \cite{preskill2018quantum}, however over the long term quantum error correction will be required to ensure that large quantum programs can execute with high success probability. Currently, the leading error correction protocol is known as the surface code \cite{dennis2002topological,FowlerSurface}, which benefits from low overheads in terms of both fabrication complexity and amount of classical processing required to perform decoding. 
 
 A common execution model of machines protected by surface code error correction requires a process called {\it magic-state distillation}. In order to perform universal computation on a surface code error corrected machine, special resources called \textit{magic states} must be prepared and interacted with qubits on the device. This process is very space and time intensive, and while much work has been performed optimizing the resource preparation circuits and protocols to make the distillation process run more efficiently internally \cite{Bravyi_magic,haah2017magic,jones2013multilevel,Fowler2013, ding2018magic}, relatively little focus has been placed upon the design of an architecture that {\it generates} and {\it distributes} these resources to a full system.
 
This study develops a realistic estimate of resource overheads of, and examines the trade-offs present in, the architecture of a system that prepares and distributes magic states. In particular, instead of using a single large factory to produce all of the magic states required for an application, the key idea of our work is to distribute this demand across several smaller factories that together produce the desired quantity. We specifically characterize these types of distributed factory systems by three parameters: the total number of magic states that can be produced per cycle, the number of smaller factories on the machine, and the number of distillation rounds that are executed by each factory. 

The primary trade-off we observe is between the number of qubits (area/space) and the amount of time (latency) spent in the system: we can design architectures that use minimal area but impose large latency overheads due to lower magic-state output rate, or we can occupy larger amounts of area dedicated to resource production aiming to maximally alleviate application latency. The two metrics, space and time, are equally important as it is easy to build small devices with more gates or large devices with few gates. This concept is closely related to the idea of ``Quantum Volume'' \cite{bishop2017quantum}, when machine noise and topologies are taken into consideration. To capture the equal importance of both of these metrics, we use a space-time product cost model in which the two metrics simply multiply together. This model has been used elsewhere in similar analysis \cite{Fowler2013,ding2018magic,paler2017fault,javadi2017optimized}.

\begin{figure}[h!]
    \centering
    \includegraphics[width=\linewidth]{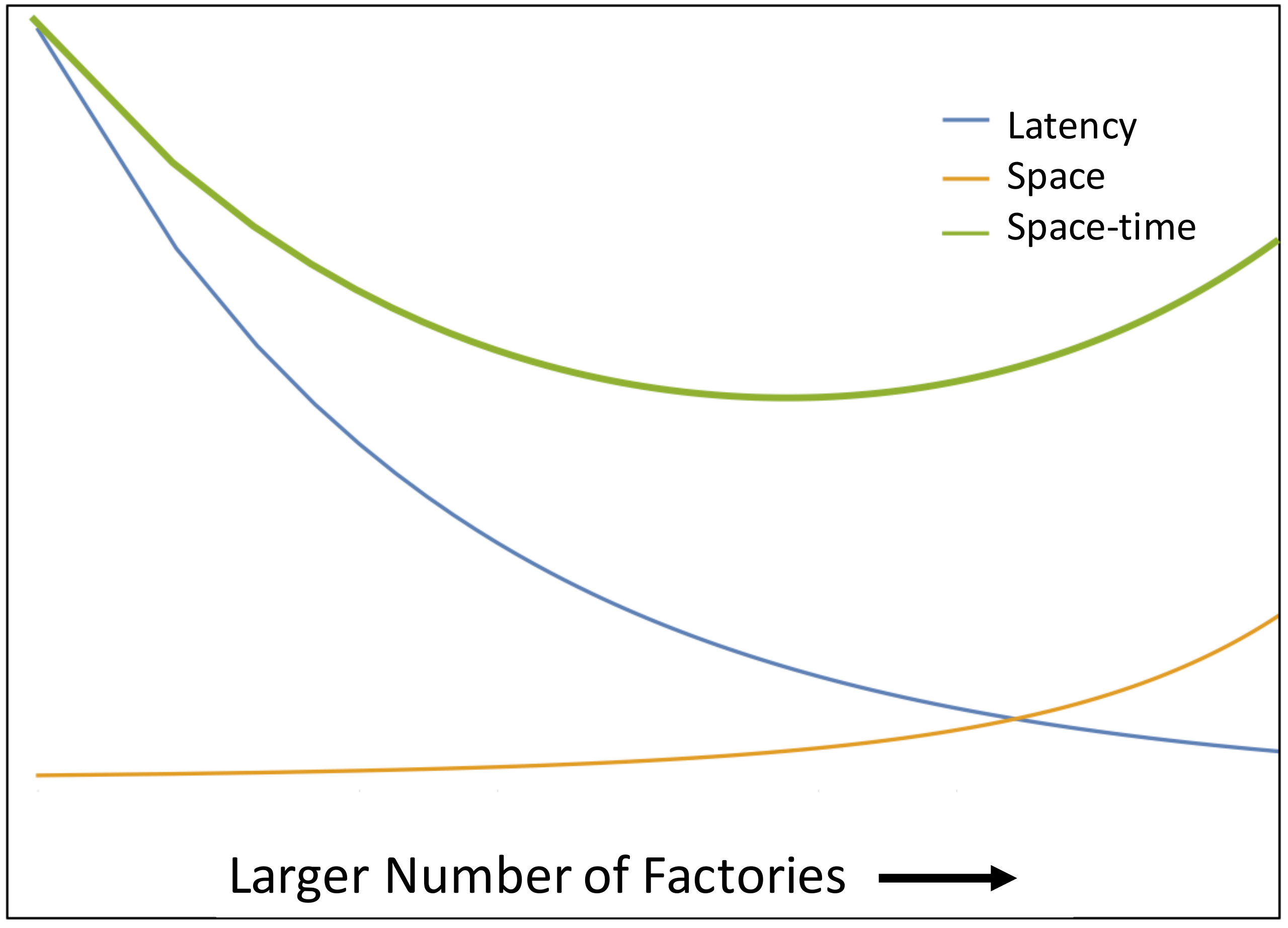}
    \caption{Space and time tradeoffs exist for distributions of resource generation factories within quantum computers. These trends are shown assuming same total factory output capacity. By explicit overhead analysis, we can discover optimal space-time volume design points.}
    \label{fig:tradeoffcartoon}
\end{figure}

Figure~\ref{fig:tradeoffcartoon} illustrates the opposing trends for space and time when we increase the magic-state production rate. Our goal is to find the ``sweet spot'' on the combined space-time curve, where the overall resource overhead is at its lowest.

In summary, this paper makes the following contributions:
\begin{enumerate}
\item We present precise resource estimates for implementing different algorithms with magic-state distillation on a surface code error corrected machine. We derive the estimates from modeling and simulating the generation and distribution of magic states to their target qubits in the computation. 
\item We quantify the space and time trade-offs of a number of architectural configurations for magic-state production, based on design parameters including the total number of factories, total number of output states these factories can produce, and the desired fidelity of the output magic states. 
\item We study different architectural designs of magic-state distillation factory, and present an algorithm that finds the configuration that minimizes the space-time volume overhead. 
\item We highlight the nontrivial interactions of factory failure rates and achievable output state fidelity, and how they affect our design decisions. We analyze the sensitivity of these optimized system configurations to fluctuations in underlying input parameters. 
\item We discover that dividing a single factory into multiple smaller distributed factories can not only reduce overall space-time volume overhead but also build more resilience into the system against factory failures and output infidelity.
\end{enumerate}

The rest of the paper is structured as follows. In Section \ref{sec:bg}, a basic background of quantum computation, error correction, magic-state distillation and the Bravyi-Haah distillation protocol, as well as the block-code state-distillation construction are described. Section \ref{sec:related} describes previous work in this area. Sections \ref{sec:area} and \ref{sec:latency} discuss important space and time characteristics of the distillation procedures that we consider, and derive and highlight scaling behaviors that impact full system overhead analysis. Section \ref{sec:tradeoffs} describes in detail how these characteristics interact, and shows how these interactions create a design space with locally optimal design points. Section \ref{sec:method} details the system configurations we model, describes a novel procedure for discovering the optimal design points, and discusses the simulation techniques used to validate our model derivations. Section \ref{sec:results} shows our results and the explains the impacts of optimizing these designs. Sections \ref{sec:conclusion} and \ref{sec:future} conclude and discuss ideas to be pursued as future work.

\section{Background}\label{sec:bg}

\subsection{Quantum Computation}
\label{subsec:qc}

The idea of quantum computation is to use quantum mechanics to manipulate information stored in two-level physical systems called quantum bits (qubits). In contrast to a bit in a classical machine, each qubit can occupy two logical states, denoted as $\ket{0}$ and $\ket{1}$, as well as a linear combination (superposition) of them, which can be written as  $\ket{\psi} = \alpha \ket{0} + \beta \ket{1} $, where $\alpha, \beta$ are complex coefficients satisfying $|\alpha|^2 + |\beta|^2 = 1$.

It is sometimes useful to visualize the state of a single qubit as a vector on the bloch sphere \cite{bloch1946nuclear,MikenIke}, as we can rewrite the state $\ket{\psi}$ in its spherical coordinates as $\ket{\psi} = \cos{(\theta/2)}\ket{0} + \exp{(i\phi)}\sin{(\theta/2)}\ket{1}$. Any operations (called quantum logic gates) performed on single qubit can thus be regarded as rotations by an angle $\varphi$ along some axis $\hat{n}$, denoted as $R_{\hat{n}}(\varphi)$. In this paper we will focus on some quantum gates that are commonly used in algorithms, such as the Pauli-X gate ($X \equiv R_x(\pi)$), Pauli-Z gate ($Z \equiv R_z(\pi)$), Hadamard gate ($H \equiv R_x(\pi)R_y(\pi/2)$), S gate ($S \equiv R_z(\pi/2)$), and T gate ($T \equiv R_z(\pi/4)$). For multi-qubit operations, we will consider the most common two-qubit gate called controlled-NOT (CNOT). It has been shown \cite{Barenco} that the above mentioned operations form a \emph{universal} gate set, which implies that any quantum operations can be decomposed as a sequence of the above gates.  


As quantum logic gates require extremely precise control over the states of the qubits during execution, a slight perturbation of the quantum state or a minor imprecision in the quantum operation could potentially result in performance loss and, in many cases, failure to obtain the correct outcomes. In order to maintain the advantage that quantum computation offers while balancing the fragility of quantum states, quantum error correction codes (QECC) are utilized to procedurally encode and protect quantum states undergoing a computation. One of the most prominent quantum error correcting codes today is the surface code \cite{dennis2002topological,FowlerSurface}.

\subsection{Surface Code}\label{subsec:surface}
In a typical surface code implementation, physical qubits form a set of two-dimensional rectangular arrays (of logical qubits), each of which performs a series of operations only with its nearest neighbors. A logical qubit, under this construction, is comprised of a tile of physical qubits, and these tiles interact with each other differently according to different logical operations. These interactions on the grid create the potential for communication-imposed latency, as routing and logical qubit motion on the lattice must be accomplished.

An important parameter of the surface code is the \textit{code distance} $d$. Larger code distance means a larger tile for each logical qubit. The precise number of physical qubits required in each tile also depends on the underlying surface code implementation. Most common implementations assume a logical qubit of distance $d$ requires $\sim d^2$ physical qubits \cite{FowlerSurface, latticesurgery}. Code distance also determines how well we can protect a logial qubit. The logical error rate $P_L$ of a logical qubit decays exponentially in $d$. More precisely:
\begin{align}
    P_L \sim d(100\epsilon_{in})^{\frac{d+1}{2}}\label{eq:etarg}
\end{align}
where $\epsilon_{in}$ is the underlying physical error rate of a system \cite{Fowler2013}.


In particular, this work will focus on two relatively expensive operations on surface code, namely the logical CNOT gate and the logical T gate. Our overhead analysis will hold regardless of the underlying technology, e.g. superconducting or ion-trap implementations. Earlier work \cite{javadi2017optimized} has also performed such analysis with technology-independent frameworks. Firstly, a logical CNOT between two qubits can be expensive, because the two logical qubits can be located far apart on the lattice and long-distance interaction is achieved by the \emph{topological defect braiding} methodology. Secondly, a logical T gate can also be costly because it requires some ancillary state to be procedurally prepared in advance, called the \emph{magic-state distillation}. 

\subsubsection{CNOT Braiding}
A \textit{braid} is a path in the surface code lattice, or an area where the error correction mechanisms have been temporarily disabled and where no other operations are allowed to use. In other words, braids are not allowed to cross. A logical qubit can be entangled with another if the braid pathway encloses both qubits, where enclosing means extending a pathway from source qubit to target qubit and then contracting back via a (possibly different) pathway. It is important to note that these paths can extend up to arbitrary length in constant time, simply by disabling all area covered by the path in the same cycle. Furthermore, each path must remain open for a constant number of surface code cycles to establish fault tolerance. More precisely, one CNOT braid takes $T_{cnot} = 2d+2$ cycles to be performed fault tolerantly \cite{FowlerSurface,javadi2017optimized}.

\subsubsection{T Magic-States}
\label{subsec:magic}

Now T (and S) gates, as described earlier, are necessary for universal quantum computation, and yet are very costly to implement on the surface code. For simplicity of analysis, we assume all S gates will be decomposed into two T gates, because of their rotation angle relationship. This is potentially an overestimate of the actual gate requirements, as it is also possible to perform an S gate via separate distillation of a different type of magic state. We are also aware of another surface code implementation that allows for S gate to be executed without distillation \cite{litinski2018lattice}. These techniques have different architectural implications which are outside the scope of the analysis of this work. 

To execute these gates, an ancillary logical qubit must be first prepared into a special state, known as the {\em magic state}~\cite{magic_states}. Once prepared, this magic-state is to be interacted with the target qubit as in \cite{FowlerSurface}, via a probabilistic circuit involving the magic state and between 1 or 3 CNOT braids, each with probability $1/2$. The extra 2 CNOTs are required to perform a corrective S gate in the case that the probabilistic circuit fails, which we assume to be consisting of 2 CNOT braids. This circuit is called the state injection circuit. We can therefore write the expected latency of a T gate as 
\begin{align}
    \mathbb{E}[T_t] &= T_{cnot} + \frac{1}{2}(2*T_{cnot}) = 4d+4\label{eq:tgatetime}
\end{align}
where we use $T_t$ to denote latency of a T gate and $T_{CNOT}$ as latency of a CNOT gate.

Since the task of preparing these states is a repetitive process, it has been proposed that an efficient design would dedicate specialized regions of the architecture to their preparation~\cite{steane1997space,Jones}. These {\em magic-state factories} are responsible for creating a steady supply of low-error magic states. The error in each produced state is minimized through a process called {\em distillation}~\cite{Bravyi_magic}, which we will introduce in detail in section \ref{subsec:BH}.

\subsection{T-Gates in Quantum Algorithms}
\label{subsec:algs}
Among the different classes of quantum algorithms, quantum simulation and quantum chemistry applications have drawn significant attention in recent years due to the promises they show in transforming our understanding of new and complex materials, while still potentially remaining tractable in near-term intermediate-size machines \cite{montanaro2016quantum,babbush2017low,kivlichan2018quantum,whitfield2011simulation,jones2012faster}. 

The benchmark algorithms studied in this work include the \emph{Ground State Estimation} (GSE) \cite{whitfield2011simulation} of the Fe$_2$S$_2$ molecule and the \emph{Ising Model} (IM) \cite{barends2016digitized} algorithms. They are representative applications for the purpose of this study as they present very different demand characteristics for T gate magic states. A more detailed description of T gate distributions in these two algorithms can be found in section \ref{subsec:program}. Here we list in Table \ref{tab:benchmarks} the two benchmarks alongside with some of their T gates statistics, namely the number of qubits ($n_{\text{qubits}}$), total T count ($T_{\text{count}}$), total schedule length ($L$), average T gates per time step ($T_{\text{avg}}$), standard deviation of T gates per time step ($T_{\text{std}}$), and maximum T gates per time step ($T_{\text{peak}}$). 

\begin{table}[h!]
    \centering
    \small
    \begin{tabular}{ccccccc}
        \hline\hline
        Application & $n_{\text{qubits}}$ & $T_{\text{count}}$ & $L$ & $T_{\text{avg}}$ & $T_{\text{std}}$ & $T_{\text{peak}}$  \\\hline
        IM & 500 & 9068348 & 20589 & 440 & 107 & 778 \\
        GSE & 5 & 775522 & 546708 & 1.419 & 1.464 & 12\\\hline
    \end{tabular}
    \caption{T gate statistics in the Ising Model (IM) and Ground State Estimation (GSE) benchmarks. For our analysis, we consider a 500-qubit spin chain in our IM simulation, and we simulate a small molecule in GSE comprised of 5 spin orbital states. The reason $T_{\text{peak}}$ for IM can be more than the number of qubits is because in this calculation every S gate in the application has been decomposed into 2 T gates.}
    \label{tab:benchmarks}
\end{table}

The Ising Model and Ground State Estimation applications, and others in the same application class, have a predictable structure. Contemporary methods to simulate quantum mechanical systems employ Trotter decomposition \cite{trotter1959product} to digitize the simulation, which involves large numbers of structurally identical Jordan-Wigner Transformation circuits \cite{batista2001generalized}, each of which involves a series of CNOT gates (called the ``CNOT staircase") followed by a controlled rotation operation. This arbitrary-angle rotation will often be decomposed to sequences of H, S, and T operations in a procedure called gate synthesis \cite{ross2014optimal}.

Take as an example finding molecular ground state energies of the molecule Fe$_2$S$_2$ requires approximately $10^4$ Trotter steps for ``sufficient" accuracy, each comprised of $7.4 \times 10^6$ rotations \cite{wecker2014gate}. Each of these controlled rotations can be decomposed to sufficient accuracy using approximately 50 T gates per rotation \cite{kliuchnikov2012fast}. All of this can amount to a total number of T gates of order $10^{12}$, which is also the number of prepared magic-states needed. In these types of applications, magic-state distillation will be responsible for between $50\% - 99\%$ of the resource costs when executing an error-corrected computation \cite{ding2018magic}. Because of this, the number of T gates present in an algorithm is often used as a metric for assessing the quality of a solution~\cite{Selinger:2013aa,amy2014polynomial}.


\subsection{Bravyi-Haah Distillation Protocol}
\label{subsec:BH}
In order to execute T gates fault tolerantly, an interaction is required between a target logical qubit and an ancillary magic state qubit. The fidelity of the operation is then tied to the fidelity of the magic state qubit itself, which requires that magic states are able to be reliably produced at high fidelity. This is achieved through procedures known as distillation protocols.

Distillation protocols are circuits that accept as input a number of potentially faulty raw magic states ($n$) and output a smaller number of higher fidelity magic states ($k$). The input-output ratio $n \rightarrow k$ is generally used to assess the efficiency of a protocol. 
Because many distillation protocols are extremely resource-intensive, a key design issue of quantum architectures is to optimize them.


In this work we restrict our focus to a popular low-overhead distillation protocol known as the Bravyi-Haah distillation protocol that has received much attention in the field recently \cite{jones2013multilevel, Fowler2013, campbell}. Here we describe in detail the process for preparing and distilling the magic-states. Bravyi-Haah state distillation circuits \cite{Bravyi_magic} take as input $3k+8$ low-fidelity states, and output $k$ higher fidelity magic-states, and thus are denoted as the $3k+8\rightarrow k$ protocol. Notably, if the raw input (injected) states are characterized by error rate $\epsilon_{\text{inject}}$ (which could be different from the physical input error rate $\epsilon_{\text{in}}$ as in equation \ref{eq:etarg} depending on hardware implementations), the output state fidelity is improved with this procedure to:
\begin{equation}\label{eq:bherror}
    \epsilon_{\text{output}} = (1+3k)\epsilon_{\text{inject}}^2, 
\end{equation}
or in other words, a second-order suppression of error.

This imposes a tolerance threshold on the underlying input error rate that can be precisely written as: 
    \begin{align}
    \epsilon_{\text{thresh}} &\approx \frac{1}{3k+1}\label{eq:thresh}
    \end{align}
because when $\epsilon_{\text{inject}} \ge \epsilon_{\text{thresh}}$, the output error rate is no better than where we started before distillation.

Moreover, this process is imperfect. For any given implementation of this circuitry, the true yield could be lower than expected. The success probability of the protocol that attempts to output $k$ high fidelity states is, to the highest order, given by:
\begin{equation}
P_{\text{success}} \approx 1-(8+3k)\epsilon_{\text{inject}}\label{eq:bhyield}.
\end{equation}

In performing a rigorous full system overhead analysis, these effects will become extremely significant.
\subsection{Block Codes}
\label{subsec:block}
In certain types of applications, the second-order error suppression achieved by single round of Bravyi-Haah distillation is not enough. To overcome this, multiple rounds (also referred to as \emph{levels} in our work) of the distillation protocol can be concatenated to obtain higher and higher output state fidelity. 

\begin{figure}[t!]
    \centering
    \includegraphics[width=\linewidth]{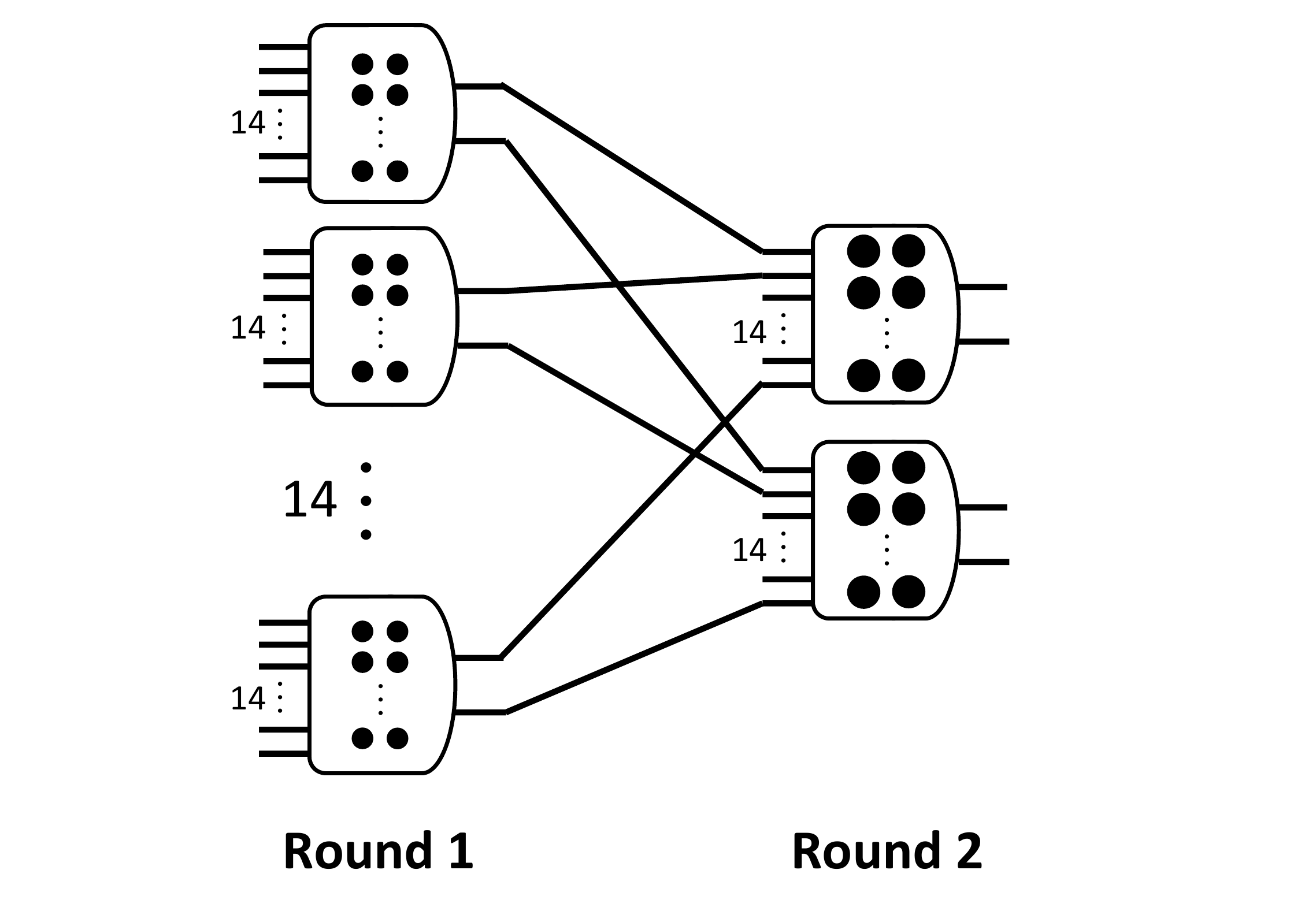}
    \caption{The recursive structure of the block code protocol. Each block represents a module for Bravyi-Haah $(3k+8)\rightarrow k$ protocol, and lines indicate the magic-state qubits being distilled, and dots indicates the extra $3k+6$ ancillary qubits used, totaling to $6k+14$. This figure shows an example of 2-level block code with $k=2$. So this protocol takes in total $(3k+8)^2=14^2$ states, and outputs $k^2=4$ states with higher fidelity. The qubits (dots) in round 2 are drawn at bigger size, indicating the larger code distance $d$ required to encode the logical qubits, as they have lower error rate than in the previous round \cite{campbell}.}
    \label{fig:block_pic}
\end{figure}

To ensure successful execution of a program, systems must be able to perform all of the gates in the computation with an expected value of logical gate error rate less than 1. So the success probability desired for a specific application ($P_s$) relates to the required logical error rate per gate $P_L$ as follows:
\begin{align}
P_L \leq \frac{P_s}{N_{\text{gates}}}
\end{align}
where $N_{\text{gates}}$ is the number of logical gates in the computation. $P_L$ therefore sets a bound on the fidelity of generated magic states. Many circuits contain of order $10^{10}$ logical gates or more \cite{wecker2014gate}, while physical error rates may scale as poorly as $10^{-3}$ \cite{Fowler2013}. In these cases, clearly squaring the input error rate will not achieve the required logical error rate to execute the program. Instead, we can \textit{recursively} apply the Bravyi-Haah circuit $\ell$ times, with permutations of the intermediate output states in between distillation rounds. Throughout this work, we use the terminology ``round'' and ``level'' to both refer to a single iteration of the Bravyi-Haah distillation protocol within a factory. Constructing high fidelity states in this fashion is known as Block Code State Distillation \cite{jones2013multilevel}. As shown in Figure \ref{fig:block_pic}, realizing Bravyi-Haah block code protocols would require $6k+14$ total logical qubits \cite{campbell}.

\subsubsection{Magic-State Factory Error and Yield Scaling}
To perform a rigorous full system overhead analysis, it is necessary to quantify the behavior of multi-level block code factories in terms of output state fidelity and production rate. By construction, the error rate of the produced magic-states will be squared after each round. So the final output states error rate after $\ell$ rounds of distillation will be $\sim \epsilon_{\text{inject}}^{2^{\ell}}$. 

Since the output states from the previous round will be fed into the next round, the success probability of a distillation module at round $r$ depends on the output error rate of the previous round $\epsilon_{r-1}$, i.e. $P_{\text{success}}^{(r)} = 1-(3k+8)\epsilon_{r-1}$. The success probability for the entire $\ell$-level factory will be explicitly derived later in Section \ref{sec:area}.

\subsubsection{Magic-State Factory Area Scaling}

Within any particular round $r$ of an $\ell$-round magic-state factory (where $1 \le r \le \ell$), the required number of \emph{physical} qubits defines the space occupied by the factory during that round. However, we will often use \emph{logical} qubit as unit area, since translating to physical qubits will simply pick up a $d_r^2$ multiplicative factor as shown in section \ref{subsec:surface}. 

In general, any particular round requires several \text{modules} each comprised of several distillation protocol circuits. A generic $n\rightarrow k$ protocol, under a $\ell$-level block code construction, will need a total number of protocols as follows:
\begin{align}\label{eq:num_modules}
N_{\text{distill}} = \sum_{r=1}^\ell N_r = \sum_{r=1}^\ell k^{r-1} n^{\ell - r}
\end{align}

\subsubsection{Magic-State Factory Time Overhead}
Each round of distillation can be shown to require $11 d_r$ number of surface code cycles\cite{campbell}. Suppose $d_r$ is the code distance for round $r$ (which depends upon the input and output error rates), we arrive at the total time to execute full distillation as:
\begin{align}
T_{\text{distill}} = 11 \sum_{r=1}^\ell d_r \label{eq:distilltime}
\end{align}

A full assessment of the area and time costs under our proposed architecture designs,will be presented in more detail in Section \ref{sec:area} and Section \ref{sec:latency}. Specifically, we discuss how factory capacity, distillation rounds of each factory, and the input physical error rate all affect the output state yield rate and resulting space and time overhead.

\section{Related Work}\label{sec:related}

\begin{figure}[t]
    \centering
    \begin{subfigure}[b]{0.23\textwidth}
    \includegraphics[width=\linewidth]{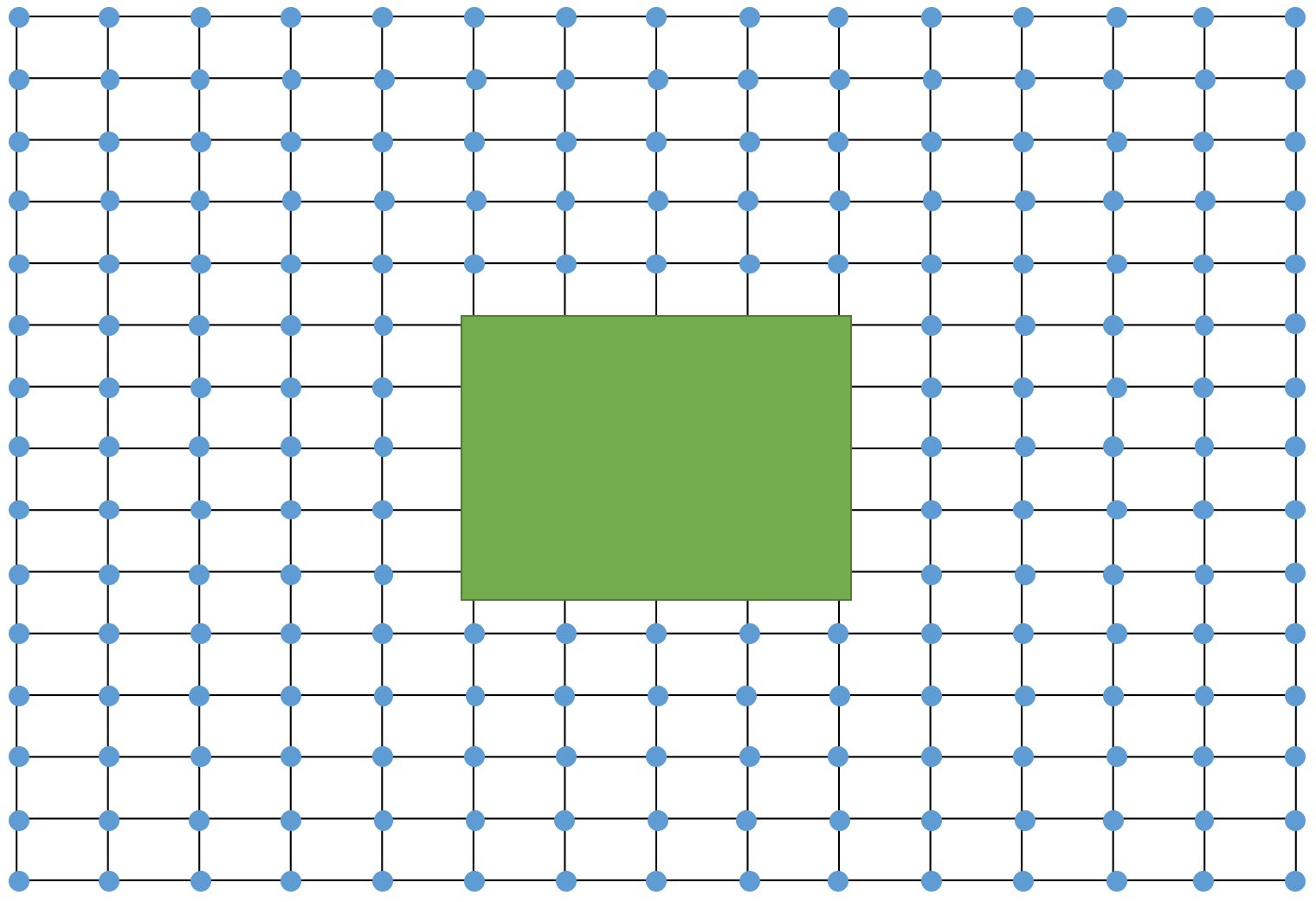}
    \caption{Single unified factory with \emph{large} capacity}
    \label{fig:unifactory}
    \end{subfigure}
    ~ 
    \begin{subfigure}[b]{0.23\textwidth}
    \includegraphics[width=\linewidth]{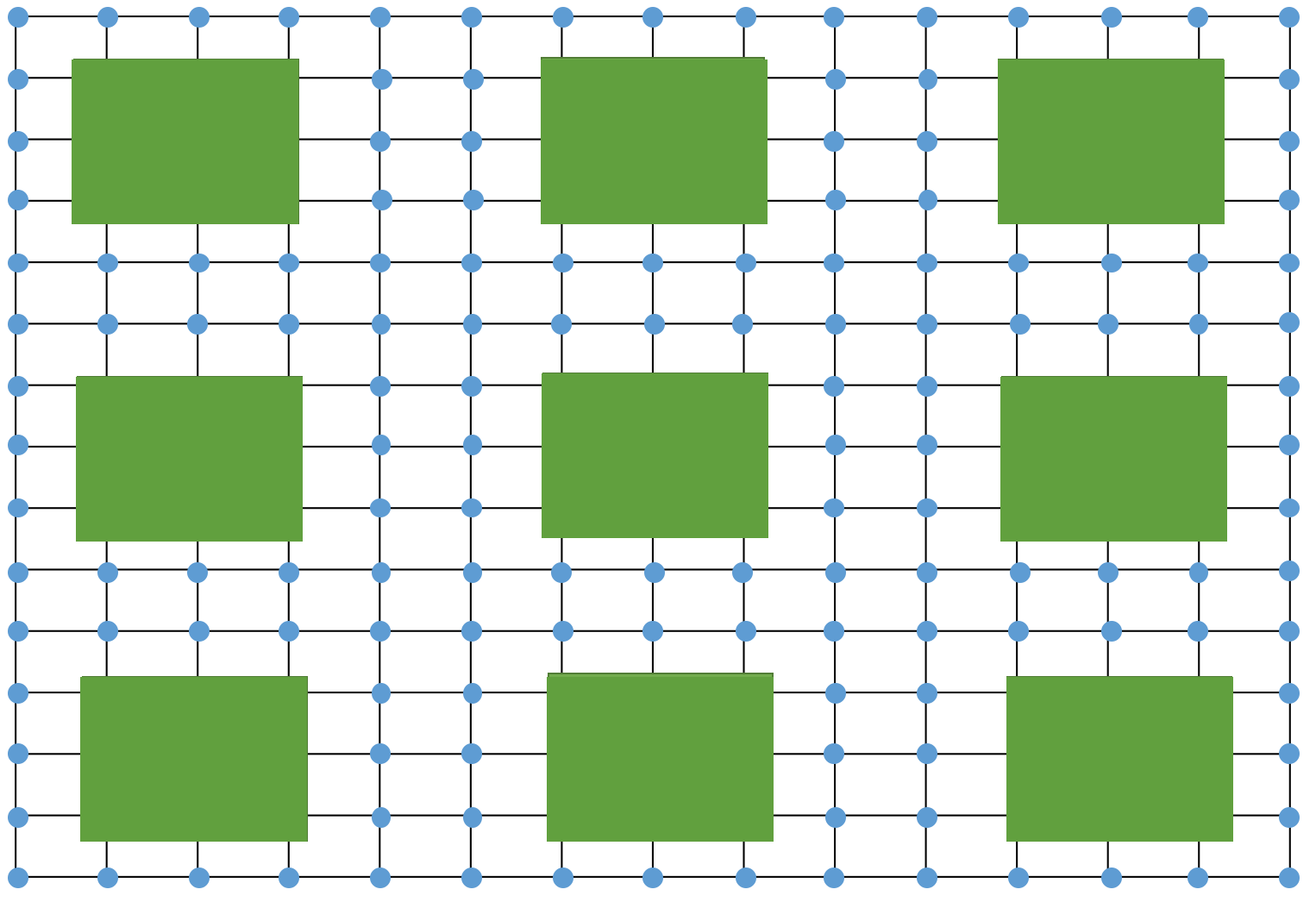}
    \caption{A number of distributed factories, each with \emph{smaller} capacity}
    \label{fig:distfactory}
    \end{subfigure}
    \caption{The concept of a unified versus distributed factory architecture, embedding factories (green blocks) within computational surface code region (blue circles).}
    \label{fig:fact_time}
\end{figure}
A number of prior work has been focused upon designing efficient magic-state distillation protocols \cite{Bravyi_magic,anwar2012qutrit,meier2012magic,magic_states}. There are also some work that aim to concatenate different protocols together to reduce the overall cost or improve output rate and fidelity \cite{jones2013multilevel,campbell2012magic}. The problem of scheduling and mapping the distillation circuit is tackled in this work \cite{ding2018magic} by taking advantages of the internal structures in the distillation protocol, and by minimizing CNOT-braid routing congestions. The aim of that work is also to more effciently implement the distillation process, which is different from ours, as we instead aim to optimize a full system architecture built around these protocols and construct factory arrangements that efficiently deliver output magic states to their intended target qubits. 

Prior work on this subject has often assumed either that magic states will be prepared offline in advance \cite{Fowler2013,Jones}, or that the production rate is set to keep up with the {\it peak} consumption rate in any given quantum application, and any excess states will be placed in a buffer \cite{campbell,van2013blueprint}. This paper operates with the different assumption that magic-state factories will be active during the computation, and states will not be able to be prepared offline or in advance. We do this to characterize the performance of the machine online, and introduce the complexity of resource state distribution throughout the machine, a problem that has been studied well in classical computing systems but has received less of a focus in this domain.

Other works closely related to architectural design optimized the ancilla production factories that operate in different error correcting codes \cite{paetznick2011fault,isailovic2008running}, or analyzed the overhead of CNOT operations which dominate other classes of applications like quantum cryptography and search optimization \cite{javadi2017optimized}. Our work focuses instead on quantum chemistry and simulation applications that are likely to represent a large proportion of quantum workloads in both the near and far term.  

\section{Factory Area Overhead}\label{sec:area}

\begin{table*}[t]
\small
\centering
\begin{tabular}{cl||cl}
\hline\hline
Parameter    &  Descriptions &   Parameter  &  Descriptions  \\\hline
$K$ & Factory total capacity  & $n$  & Number of input states in distillation protocol \\
$X$ & Number of distributed factories & $k$  & Number of output states in distillation protocol \\
$\ell$ & Block-code levels of a factory & $N_{\text{r}}$ & Number of protocols at round $r$ under block-code\\
$r$ & Distillation round, $1 \le r \le \ell$ & $K_{\text{output}}$ & Number of effective output magic-states due to yield rate\\
$d$ & Surface code distance & $T_{\text{distill}}$ & Time to execute one full iteration of distillation \\
$P_{\text{s}}, P_{\text{success}}$ & Target success probability & $T_\text{t}$ & Time to deliver magic-state to target qubit\\
$P_{L}$ & Logical fidelity & $n_{\text{distill}}$ & Distillation iterations to support one timestep of a program\\
$\epsilon_{\text{inject}}$ & Physical error rate of raw magic-state & $A_{\text{factory}}$ & Total area of factories (in physical qubits)\\
$\epsilon_{\text{in/target/r}}$ & Physical error rate at input, at output, or at round $r$  &  & \\
\hline\hline
\end{tabular}
\caption{List of system parameters involved in the analysis and the optimization procedure.}
\end{table*}

To describe a magic-state distillation factory, we first make a distinction between a factory \textit{cycle} and a distillation \textit{round} or \textit{level}. A distillation round refers to one iteration of the distillation protocol, a subroutine that is repeated $\ell$ times for a particular factory. A cycle refers to the total time required for the factory to operate completely, taking $n$ input states and creating $k^{\ell}$ output states. All $\ell$ distillation rounds are performed during a cycle. 

{\bf A magic-state distillation factory architecture can be characterized by three parameters:} the total number of magic states that can be produced per distillation cycle $K$, number of factories on the lattice $X$, and the total number of distillation rounds that are performed per cycle $\ell$. For simplicity, we assume uniform designs where all $K$ output states are to be divided equally into $X$ factories, all of which operate with $\ell$ rounds of distillation. We now analyze the relationships presented in Section \ref{sec:bg} to derive full factory scaling behaviors with respect to these architectural design variables. These behaviors interact non-trivially, and lead to space-time resource consumption functions that show optimal design points. 

\begin{figure*}[h!]
    \centering
    \begin{subfigure}[b]{0.4\textwidth}
        \includegraphics[trim=0 0 0 0,
        width=\linewidth]{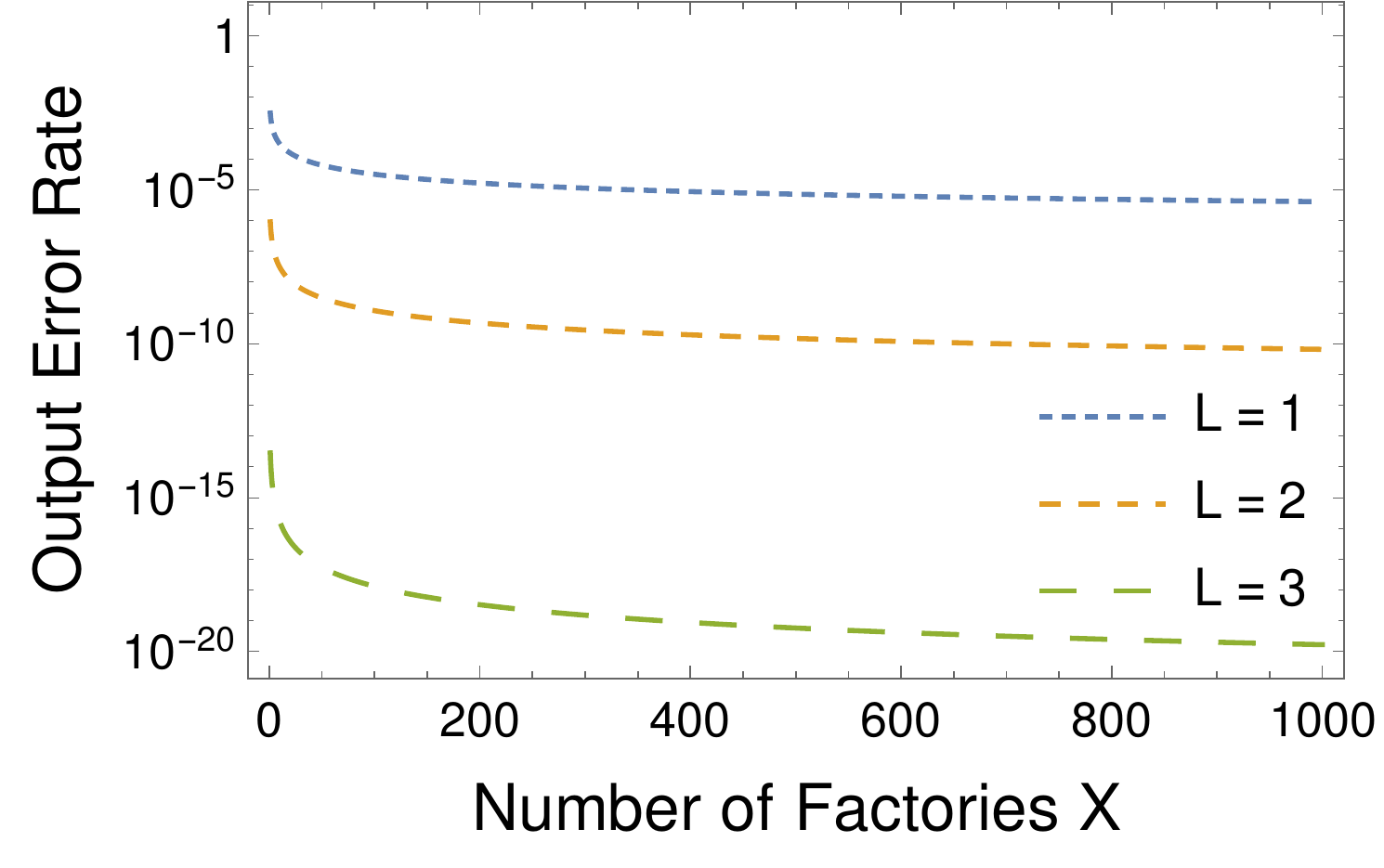}
        \caption{Error rate attainable by number of factories}
        \label{fig:yield}
    \end{subfigure}
    \begin{subfigure}[b]{0.4\textwidth}
    \includegraphics[width=\linewidth]{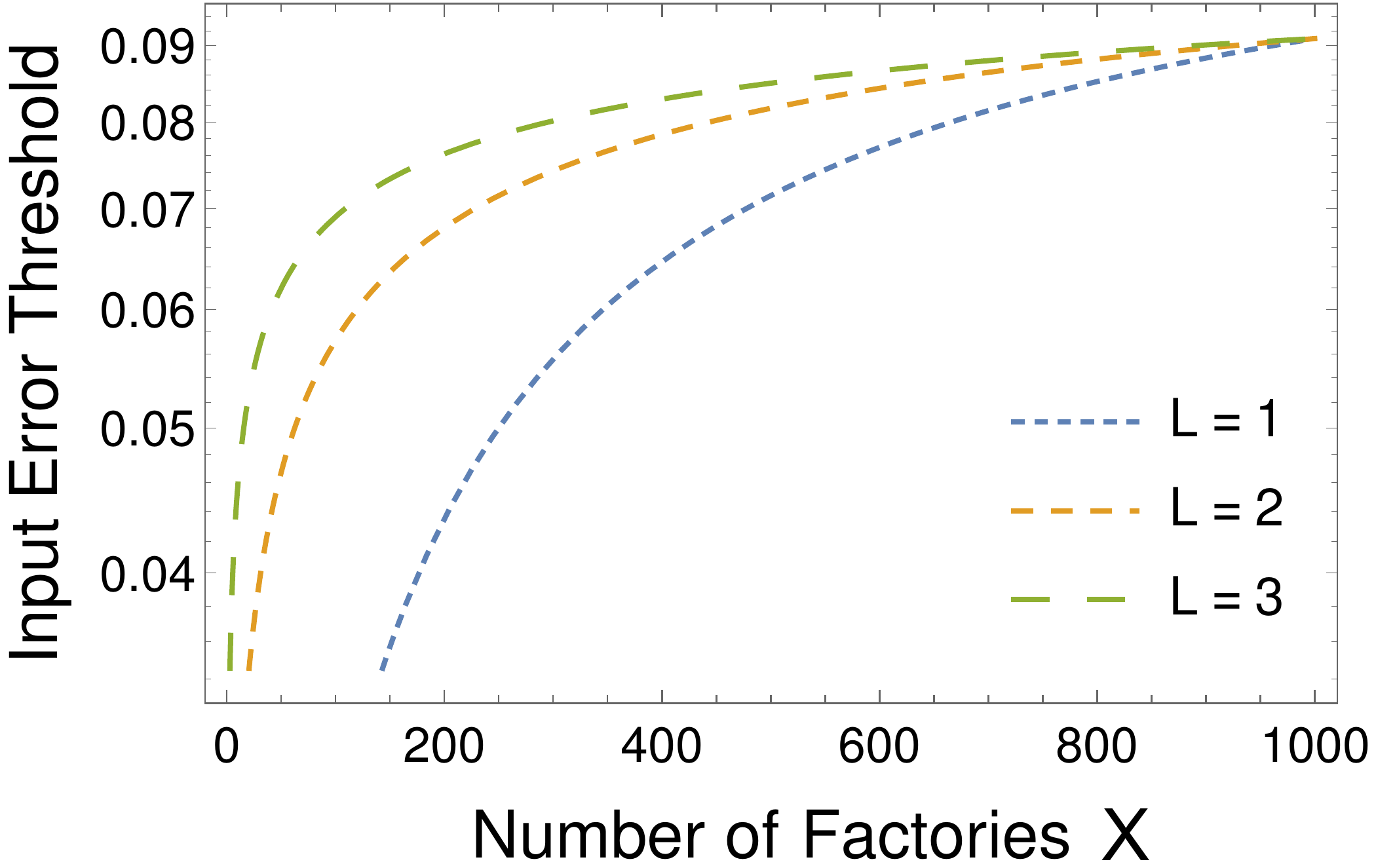}
    \caption{Error rate tolerable by number of factories}
    \label{fig:xsweeperrorthresh}
    \end{subfigure}  
    \begin{subfigure}[b]{0.4\textwidth}
        \includegraphics[       width=\linewidth]{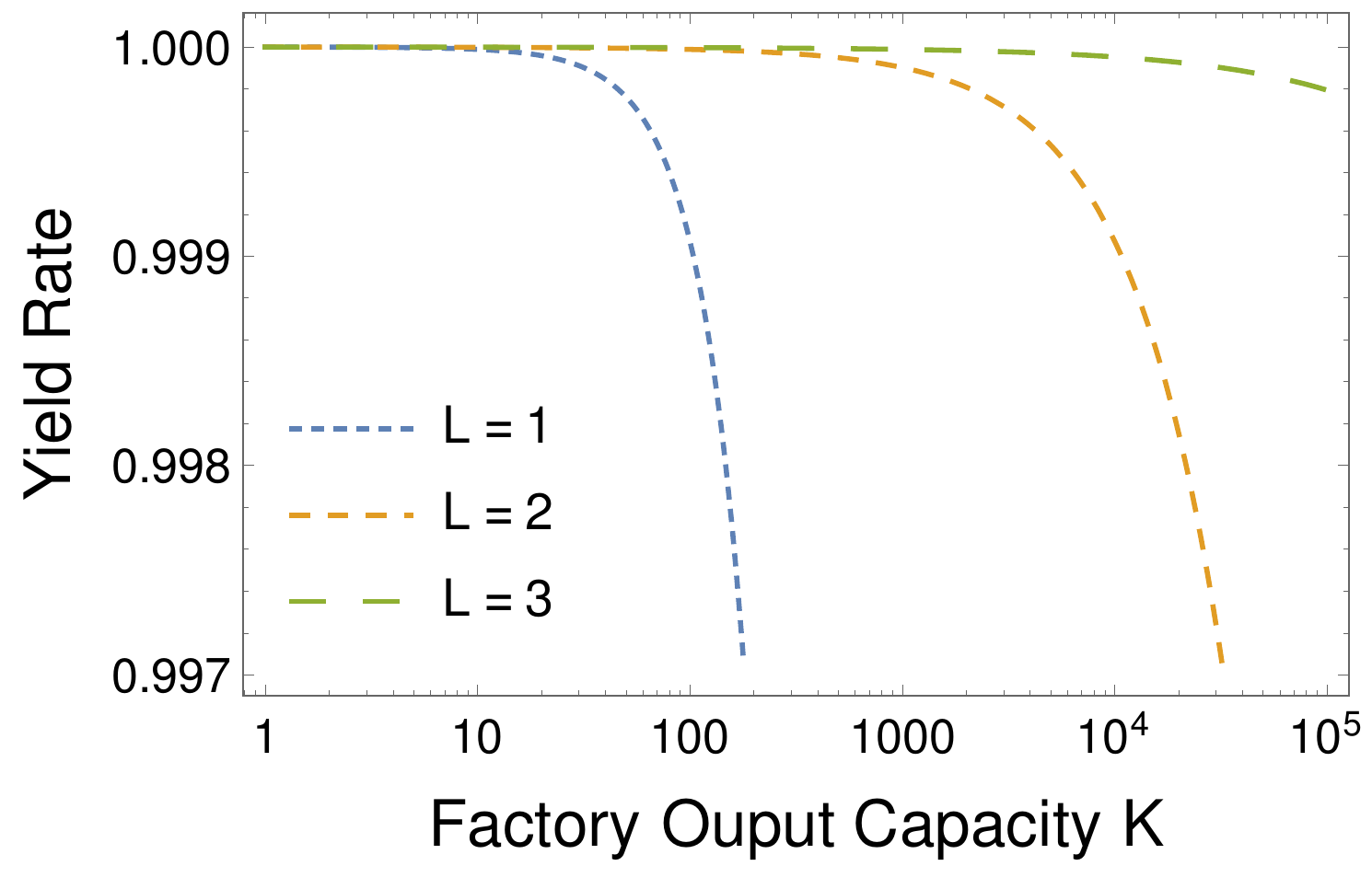}
        \caption{Yield rate of $L$-level factory with capacity $K$}
        \label{fig:yield}
    \end{subfigure}
    \begin{subfigure}[b]{0.4\textwidth}
        \includegraphics[width=\linewidth]{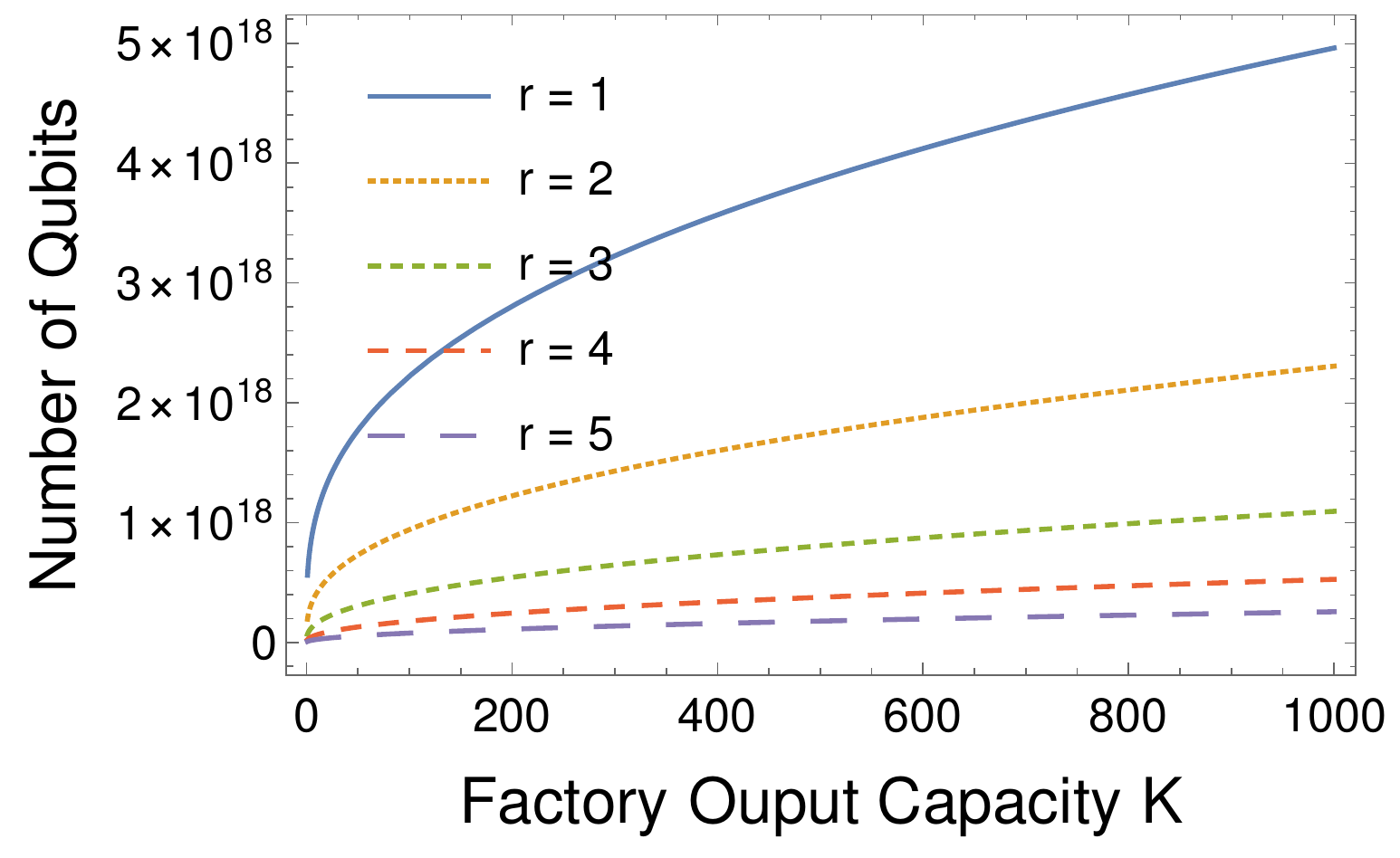}
        \caption{Area scaling within each round of a 5-level factory}
        \label{fig:area_law}
    \end{subfigure}
    \caption{(a) Higher fidelity output states are achievable with increasing number of factories at a fixed output capacity. (b) Increasing the number of factories in an architecture allows for higher tolerance of input physical error rates. (c) Increasing factory output capacity puts pressure on the factory yield rate, and increasing the number of levels pushes the yield dropoff point. (d) Maximum area to support multi-level factory is required of the lowest level of the factory, all higher levels require less area support.}
    \label{fig:factorieserrorsyields}
    \label{fig:factorycharacteristics}
\end{figure*}

\subsection{Role of Fidelity and Yield in Area Overhead}

First we examine the fidelity of the produced magic-states that is attainable with a given factory configuration, along with expected number of states that will in fact be made available. Once again, we use the terminology ``round'' and ``level'' to both refer to a single iteration of the Bravyi-Haah distillation protocol within a factory. Applying the block code error scaling relationship described by equation \ref{eq:bherror} recursively, as the number of total rounds ($\ell$) of a magic-state factory increases, the output error rates attainable scale double-exponentially with the total number of rounds in a factory: $\ell$. In fact, for a given round $r$ (between $1$ and $\ell$) of a factory, the explicit form of the output error rate can be written by directly applying $r$ copies of equation \ref{eq:bherror}:
\begin{align}
\epsilon_r &= (1+3(K/X)^{\sfrac{1}{\ell}})^{2^r-1}\epsilon_{\text{inject}}^{2^r}\label{eq:errorround}
\end{align}
where $(K/X)$ denotes the capacity of each factory on a lattice. 

The yield rate of a particular factory can be expressed as a product of the yield rate functions describing each individual round, as in equation \ref{eq:bhyield}. The effective output capacity can be written as the product of the success probabilities of all $\ell$ rounds of a factory as:
\begin{align}
K_{\text{output}} &= K \cdot \prod_{r=1}^{\ell}\bigg[(1-(3(K/X)^{\sfrac{1}{\ell}}+8)\epsilon_{r-1}\bigg]\label{eq:koutput}
\end{align}

Here $K_{\text{output}}$ refers to the realized number of produced states after adjusting for yield effects, while $K$ refers to the desired or specified number of output states. Equation \ref{eq:koutput} actually imposes a \textit{yield threshold} on the system. For a given $K$, $X$, and $\ell$, a system will have a maximum error rate which, if exceeded, will cause the factory to malfunction and stop producing states reliably. This threshold can be seen by examining the product term, and noting that yield must be positive in order to produce any states. The terms in the sequence of equation \ref{eq:koutput} are decreasing in magnitude, so the threshold is determined by the leading term which requires: $1-(3(K/X)^{1/\ell} + 8)\epsilon_{\text{inject}} > 0$, and thus:
\begin{align}
    \epsilon_{\text{thresh}} < \frac{1}{3(K/X)^{1/\ell}+8}\label{eq:yieldthresh}
\end{align}

Figure \ref{fig:yield} shows the yield rate scaling behavior of single factories of consisting of $\ell=1,2,3$ with fixed $X=1$. In order to reliably produce some fixed amount of states, the yield effects determine the required number of rounds of distillation that must be performed. On the other side, any given number of distillation rounds has a maximum output capacity $K$ for which the expected number of produced states becomes vanishingly small. Increasing the number of distillation rounds will increase the maximum supportable factory capacity. 

    
\subsection{Full Area Costs}
We now use these relationships to derive the true area scaling of these factories. For all $\ell$ level factories, the area of the first round exceeds the area required for all other rounds. Using this as an upper bound, we can write the area required for a specific round explicitly in terms of physical qubits as:
\begin{align}
   A_r&=X\cdot k^{r-1}(3k+8)^{\ell-r}(6k+14)\cdot d_r^2 \label{eq:arealaw}\\
   &\leq X(3k+8)^{\ell-1}(6k+14)\cdot d_1^2
\end{align}
Where $k \equiv (K/X)^{\sfrac{1}{\ell}}$. The inequality in the last line arises due to the fact that the first round always uses the largest area by block-code construction, i.e. $A_r \leq A_1$ for all $1 \le r \le \ell$. Here we have used several relationships, namely that the total number of protocols and modules scales as in equation \ref{eq:num_modules}, a single protocol requires $6k + 14$ logical qubits \cite{campbell}, and the area of a single logical surface code qubit scales as $d^2$ \cite{latticesurgery}. 

\begin{figure*}[t]
    \centering
    \begin{subfigure}[b]{0.3\textwidth}
        \includegraphics[width=\textwidth]{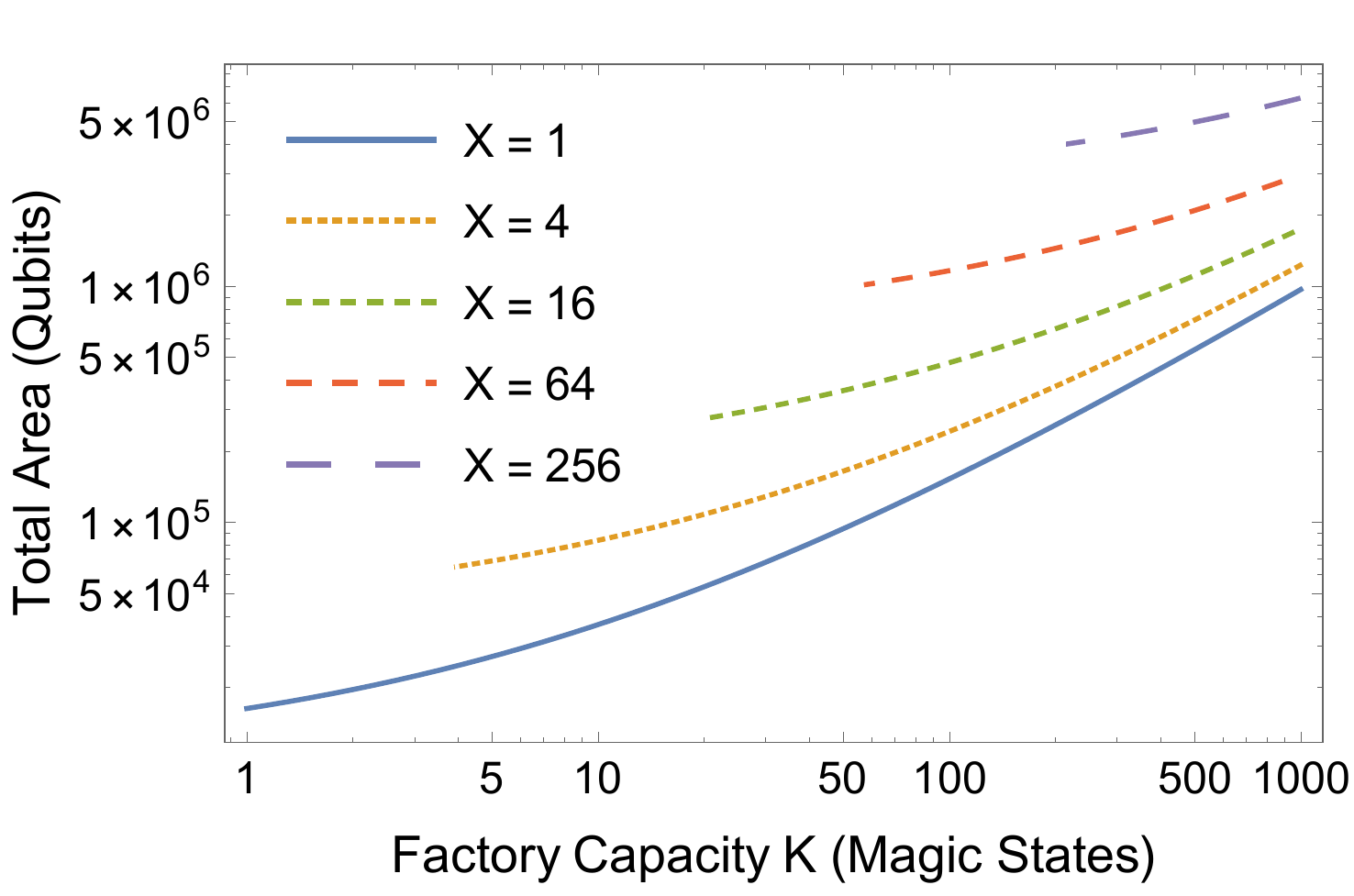}
        \caption{Area Scaling}
        \label{fig:area_k_x}
    \end{subfigure}
    ~
    \begin{subfigure}[b]{0.3\textwidth}
        \includegraphics[ width=\textwidth]{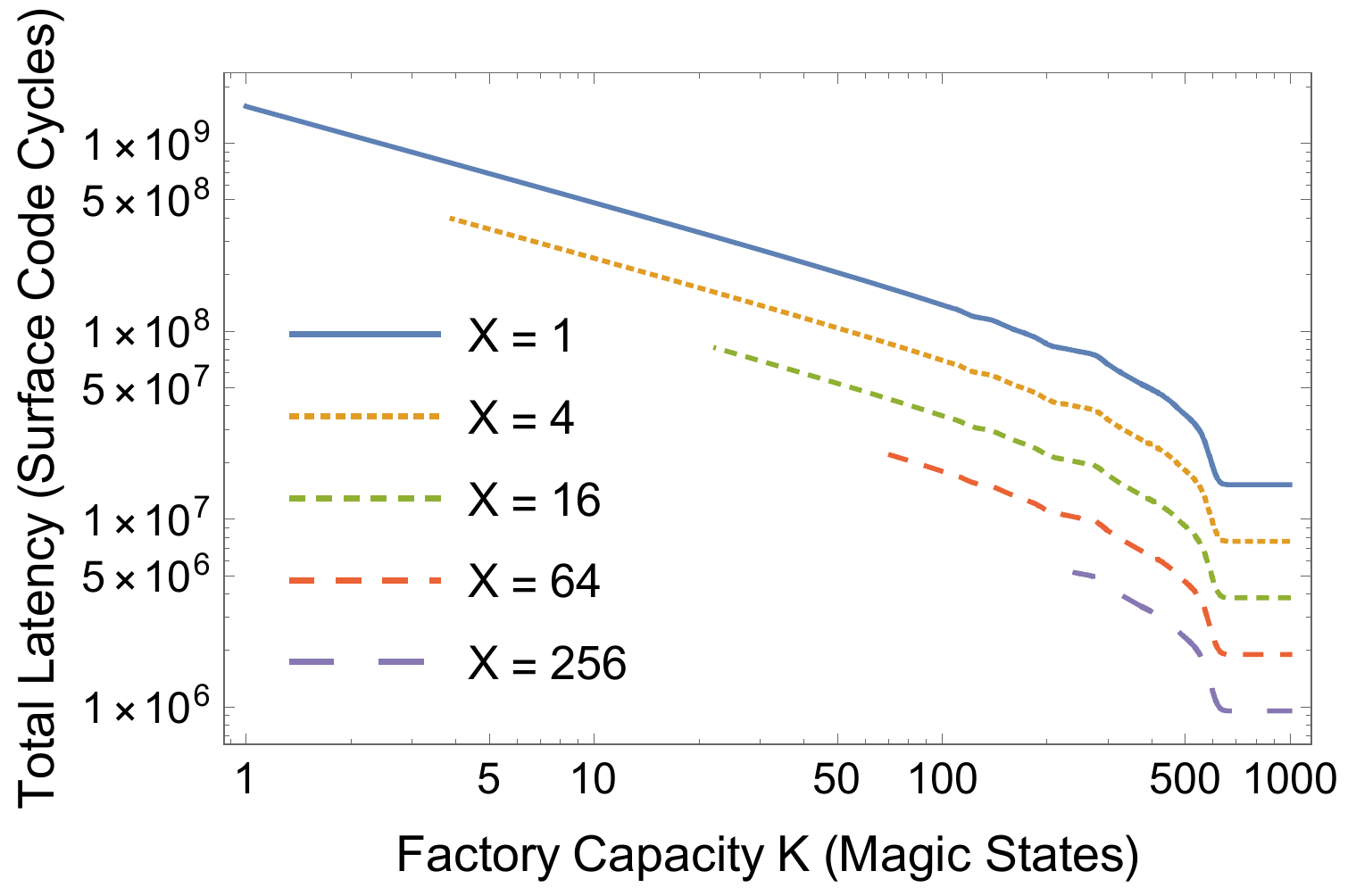}
       \caption{Latency Scaling}
      \label{fig:time_k_x}
    \end{subfigure}
    \begin{subfigure}[b]{0.3\textwidth}
        \includegraphics[ width=\textwidth]{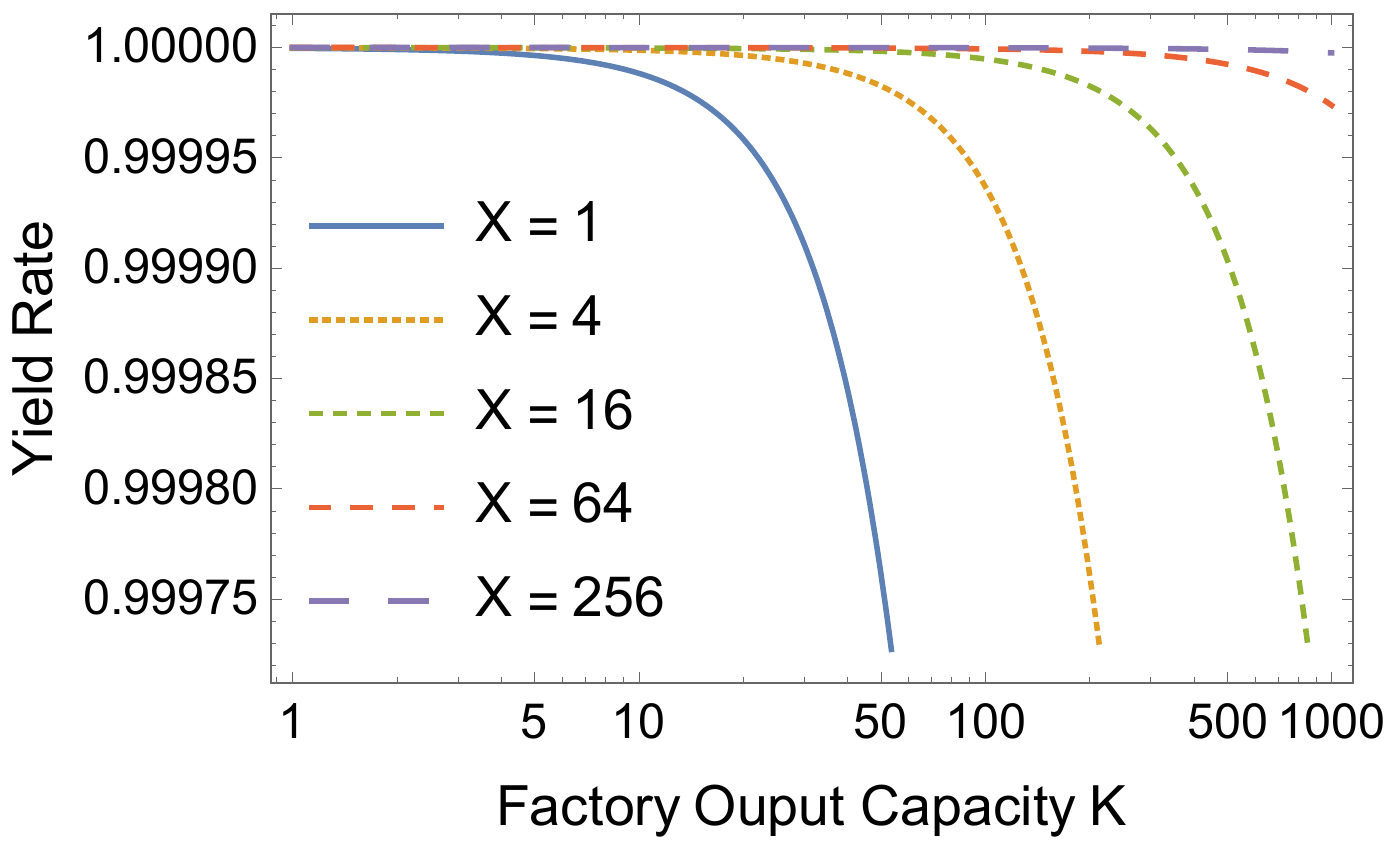}
       \caption{Yield Rate Scaling}
      \label{fig:yield_k_x}
    \end{subfigure}
    ~
        \caption{(a) Area required to implement a 2-level factory of varying numbers of factories $X$. As the distribution intensity increases, the total area increases significantly faster as factory output is scaled up. Notice that some regions are not feasible due to the constraint $K/X \le 1$. (b) Latency as it scales with factory output capacity. For factories of a fixed capacity, increasing the number of factories on the lattice reduces latency overall and speeds up application execution time, thanks to reductions in contention and congestion. The flat tails at high $K$ values are due to the fact that the capacity has exceeded the amount that a application ever demanded. (c) Yield as it scales with factory output capacity and number of factories. For a fixed capacity $K_0$, increasing the number of factories can significantly increase the success probability and yield rate of the factory.}
    \label{fig:factoriessacetime}
\end{figure*}

Although in an aggressively optimized factory design then, one could conceivably save space within the distillation procedure by utilizing the space difference between successive rounds of distillation for other computation, we will assume in this work that this cannot be done, and instead the first round area of any given factory defines the area required by that factory over the length of its entire operation, and {\it locks out} the region for distillation only. As a result, Figure \ref{fig:area_k_x} describes the scaling of factory area both by increasing output capacity and increasing the total number of factories. 

\section{Factory Latency Overhead}\label{sec:latency}
 
This section presents a systematic study of the time overhead of realizing magic-state distillation protocols. First, we will examine the characteristics of the T gate demand in our benchmark programs, by introducing the concept of the T distribution. Next, we will study the latency overhead caused by delivering magic states to wherever T gates are demanded by looking at the contention and congestion factors. Finally, we will arrive at an analytical model for the overall distillation latency integrating the information from the program distribution.

\subsection{Program Distributions}\label{subsec:program}

While the majority of the prior works on this subject have been abstracting algorithm behavior into a single number, the total T gate count, we argue that the distribution of T gate throughout a algorithm has a significant impact on the performance of the magic-state factory. For example, a program with bursty T distribution, where a large number of T gates are scheduled in a few time steps, puts significant pressure on the factory's capability of producing a large amount of high fidelity magic states quickly.

In order to quantify this behavior, we choose two quantum chemistry algorithms that represent the two extremes of T gate parallelism. On one hand, the {\it Ground State Estimation} algorithm is an application with very low T gate parallelism. An algorithm attempting to find the ground state energy of a molecule of size $m$, this application can be characterized by a series of rotations on single qubits \cite{whitfield2011simulation}. {\it Ising model}, on the other hand, is a highly parallel application demanding T gates at a much higher rate. This application simulates the interaction of an Ising spin chain, and therefore requires many parallelized operations on single qubits, along with nearest neighbor qubit interactions \cite{barends2016digitized}. To capture application characteristics, we use the ScaffCC compiler toolchain that supports application synthesis from high-level quantum algorithm to to physical qubit layouts and circuits \cite{scaffcc}. 

The majority of the time steps in Ising Model algorithm has a large number of parallel T gates with a mean T load of 440, where as Ground State Estimation has no more than 12 T gates at each time steps. As opposed to just using the single T gate count to characterize algorithms, we will from now on use the T load distribution. 

\begin{figure*}[h!]
    \centering
    \begin{subfigure}[b]{0.48\textwidth}
        \includegraphics[width=\linewidth]{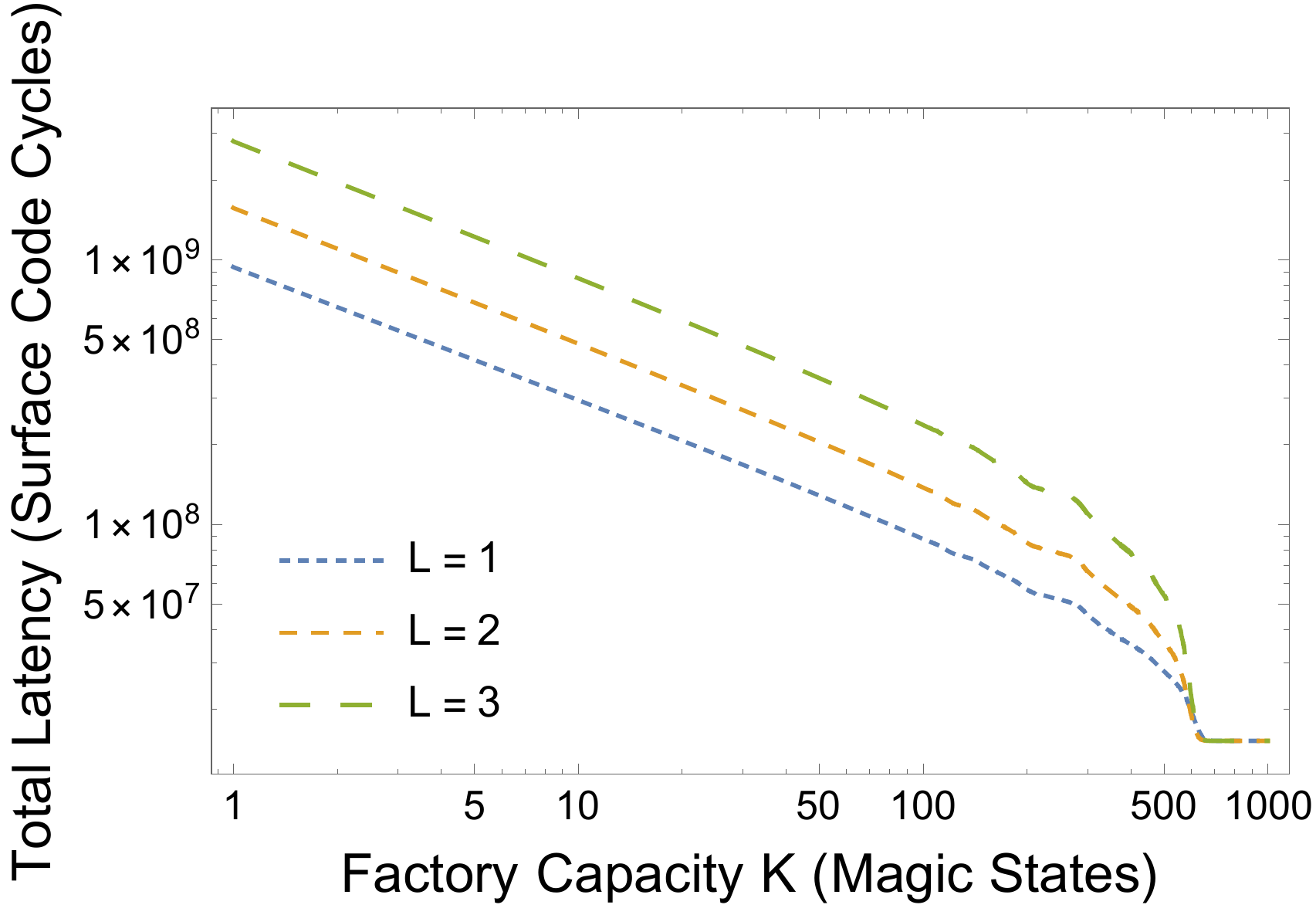}
        \caption{Ising Model}
        \label{fig:time_is}
    \end{subfigure}
    ~ 
    \begin{subfigure}[b]{0.48\textwidth}
        \includegraphics[width=\linewidth]{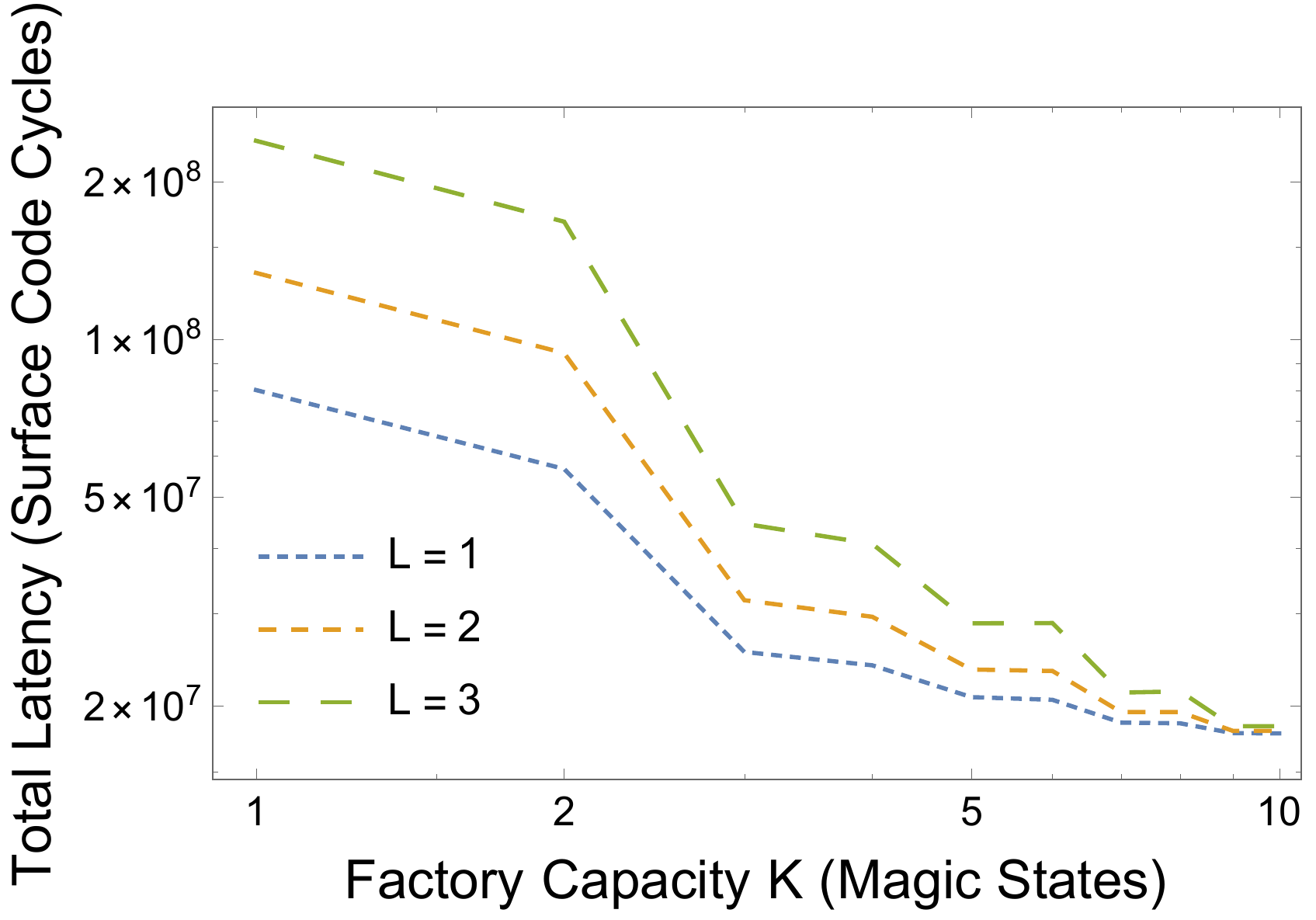}
        \caption{Ground State Estimation}
        \label{fig:time_gse}
    \end{subfigure}
    \caption{(a)-(b) Total number of surface code cycles required by Ising Model and Ground State Estimation applications. Both  figures are plotted for three different factory block-code levels, i.e. $X=1$ and $L=1, 2, \text{ and }3$. }
    \label{fig:fact_time}
\end{figure*}

\subsection{T-Gate Contention and Congestion}\label{subsubsec:contentioncongestion}
In order to fully assess the space-time volume overhead of the system, we require a low level description of how the produced magic-states are being consumed by the program. 

As discussed in the Section \ref{sec:bg}, a T gate requires braiding between the magic-state qubit in the factory and the target qubit that the T gate operates on. Now suppose our factory is able to produce $K$ high-fidelity magic states per distillation cycle, and at some time step the program requests for $t$ T gates. If we demand more than the factory could offer at once (i.e. $t > K$), then naturally only $K$ of those requests can be served, while the others would have to wait for at least another distillation cycle. So we will say that the network has {\it contention} when the demand exceeds the supply capacity. By contrast, we define network {\it congestion} to capture the latency introduced by the fact the some braids may fail to route from the target to the factory on the 2D surface code layout, due to high braiding traffic.

To estimate the overhead of network congestion, we will perform an average case analysis without committing to a particular routing algorithm. Ideally, in the contention free limit where the number of requests $t$ is less than $K$, all requests could be scheduled and executed in parallel. However, often times the requests will congest due to limitations of routing algorithms. We define a congestion \textit{factor} $C_g$that represents the total latency required to execute all of the T gate requests at any given time. 

We model congestion as a factor that scales proportional to the number of $t$ requests made at any given time, within a particular region serviced by a factory. This assumes a general topology in which a factory is placed in the center of a region, and all of the surrounding data qubits are served by this factory alone. Naturally, the center of the region is quite dense with T gate request routes. In general, for a reasonable routing algorithm, the number of routing options increases as area available increases. However, because all of the routes have their destination in the center of the region, increasing area of the region has no such effect. In fact, the \textit{distance} of a T request source from the factory increases the likelihood of congestion from a simple probabilistic argument. There may be other T requests blocking available routes, and the number of these possible requests that block pathways increases as the distance between a request and the factory increases. The combination of these effects interacts with the complexity of a routing algorithm, and results in a scaling relationship proportional to both the T request density $t$ and the maximum distance of any T request within any of these regions:
\begin{align}
    C_g \sim c\sqrt{t} 
\end{align}
for some constant $c$, depending upon the routing algorithm.

We validated this congestion model in simulation using simulation tools and compiler toolchains of \cite{javadi2017optimized}, and find that they do indeed agree. Section \ref{sec:method} discusses this in greater detail.

\subsection{Resolving T-gate Requests}\label{subsec:exec}

For any given program, characterized as a distribution $D$ of the T load, we denote $D[t]$ the number of timesteps in the program that $t$ parallel T gates are to be executed. Then the number of iterations that the factory needs to resolve the $t$ requests can be computed based on the following latency analysis. In particular, in order to maximize the utilization of the factory, we would execute as many outstanding T gate requests as possible in parallel. When the number of requests $t$ exceeds the factory yield $K$, we will need to stall the surpassed amount of requests. We denote $s = \lfloor t/K \rfloor$ the number of fully-utilized iterations. So, we are serving at full capacity for $s$ number of times, and at each time a congestion factor is being multiplied, as discussed in Section \ref{subsubsec:contentioncongestion}. It follows that the first $sK$ requests are completed in $s \sqrt{K}$ number of distillation cycles. And finally the rest $(t-sK)$ outstanding requests are then being executed in $\sqrt{t-sK}$ cycles. Notice that the time it takes to execute the T gate is typically shorter than the factory distillation cycle time. So under the buffer assumption made earlier, we can stage the execution of requests within a distillation cycle such that no data dependencies are violated, as long as there are magic-states available in the factory. The time required to produce some constant number $k$ of states is $T_{\text{distill}}$, while the time required to deliver $k$ states in parallel is $T_t\sqrt{k}$ due to network congestion. So the number of distillation cycles needed to supply a single cycle of $k$ T gate requests is given by the ratio $T_{\text{t}}\sqrt{k}/T_{\text{distill}}$. Substituting $k = K/X$ and $k = (t - sK)/X$ as described earlier, we can calculate the number of distillation iterations we need to serve $t$ T gates in a particular timestep, as:
\begin{align}
    n_{\text{distill}} = \frac{T_{\text{t}}}{T_{\text{distill}}} \cdot \Bigg(s \cdot \sqrt{\frac{K}{X}} +  \sqrt{\frac{t-sK}{X}}\Bigg)
\end{align}
where $K$ is again the yield of each iteration from Equation \ref{eq:koutput}. 

Putting it together, we obtain our final time overhead of an application:
\begin{align}
    T_{\text{total}} &= T_{\text{distill}} \cdot \Big(\sum_{t = 0}^{T_{\text{peak}}} n_{\text{distill}}\cdot D[t]\Big)
\end{align}
where $T_{\text{peak}}$ is the maximum number of parallel T gates scheduled at one timestep. Notice that this is independent of $T_{\text{distill}}$, as the distillation cycle time has been captured by the ratio $T_t/T_{\text{distill}}$. The scaling of this function is shown in Figure \ref{fig:time_k_x}, and is compared in Figure \ref{fig:fact_time} across different applications.

\section{Area and Latency Trade-offs}\label{sec:tradeoffs}

In this section, we will discuss some of the motivations of our proposed algorithm for optimizing space-time resource overhead, based on the area and latency analysis that we built up in the previous sections. 

The Bravyi-Haah protocol shows an area \textit{expansion} when a single factory is ``divided'' into many smaller factories, that is, the total area of $x$ number of factories each with some capacity $k$ is larger than the area of a factory with capacity $x\cdot k$. Figure \ref{fig:area_k_x} illustrates this trend, arising from the original area law equation \ref{eq:arealaw}.

{\bf Why do we want a distributed factory architecture?} Although it might first seem undesirable to divide a single factory into many factories due to the area expansion, there are many advantages when doing so. One such advantage is that smaller factory can produce states with higher fidelity. So, for a fixed output capacity $K$, incrementing the number of factories used to produce in total that $K$ allows for all of those $K$ states to have higher fidelity. The output error rate scales inversely with the number of factories on the lattice for a fixed output capacity $K$ as seen in Equation \ref{eq:errorround}.

This provides us with the unique ability to actually manipulate the underlying \textit{physical error rate threshold}. In particular, substitution of $K/X$ for $K$ in all of the previous equations shows that the yield threshold now also has inverse dependence upon the number of factories used.

As Figure \ref{fig:xsweeperrorthresh} shows, for a fixed output capacity and block code level $\ell$, increasing the number of factories on the lattice can greatly increase the tolerable physical error rate under which the factory architecture can operate. 

With this knowledge, we are immediately presented with architectural tradeoffs. Using the representation of programs as distributions of $T$ gate requests, any application can be characterized by a $T_{\text{peak}}$, again defined as the highest number of parallel T gate requests in any timestep of an application. For a ``surplus'' configuration, a system may set the factory output rate $K = T_{\text{peak}}$, so as to never incur any latency during the program execution. However, as the threshold in equation \ref{eq:thresh} indicates, this sets an upper bound on the tolerable input error rate $\epsilon_{\text{in}}$. With a distributed factory architecture, this provides a system parameter enabling systems to be designed that will be able to tolerate higher error rates, and still achieve the same output capacity $K$, at the expense of area as seen in the area law relationship from Figure \ref{fig:area_k_x}. Conversely, systems that are constructed with great knowledge of low underlying physical error rates may be able to reduce overall area of a surplus factory configuration by reducing the number of individual factories to a certain point. These are the tradeoffs in the design space that this work explores, and in fact we can find for representative benchmarks, configurations that are lower in capacity that can save orders of magnitude in space-time overhead overall.

\section{Evaluation Methodology}\label{sec:method}
\subsection{System Configuration}
Here we lay out all of the assumptions made about the underlying systems that we are studying. 

\begin{table}[b]
    \centering
    \small
    \begin{tabular}{c|l}
    \hline\hline
    Configuration & Description\\\hline 
        \multirow{ 3}{*}{Surplus} & One central factory that can produce enough\\
        & states to always meet the demand at each time-\\
        & step of the program as in \cite{isailovic2008running,van2013blueprint,campbell}. \\\hline
        \multirow{ 2}{*}{Singlet} & One central factory that uses minimal \\
        & area and produces only one state per cycle. \\\hline
        \multirow{2}{*}{Optimized-Unified}& One central factory that outputs an optimized \\
        & number of output states per distillation cycle\\ \hline
        \multirow{2}{*}{Optimized-Distributed}& A optimized set of factories that together \\
        & output an optimized number of output states \\ \hline
        \hline
    \end{tabular}
    \caption{List of architecture configurations explored in this work.}
    \label{tab:configurations}
\end{table}

First, we assume that the factories will be operated continuously. This means that each $T_{\text{distill}}$, the factories will produce another $K_{\text{output}}$ states. This abstracts away the time needed to deliver these states to their destinations, which would have to be performed in a real system before the next distillation iteration begins. In such real systems, we imagine an architecture that supports a limited, fixed size buffer region so that the subsequent distillation cycle will not overwrite the previously completed states. However, this is a small constant offset in time that applies to all studied designs symmetrically, so it is omitted. Because the factories are always online and producing magic states, the overall time overhead is then equal to the number of distillation cycles required to execute all the scheduled $T$ gate requests, multiplied by the time taken to perform a distillation iteration  $T_{\text{distill}}$ from Equation \ref{eq:distilltime}.

Next, we assume three different levels of uniformity in these designs: all distributed factories are laid out uniformly on the surface code lattice as in figure \ref{fig:distfactory} (i.e. they are an equal distance apart), all factories in a distributed architecture are identical (i.e. they all operate with the same parameters such as $K$ and $\ell$), and within each factory each block code round is identical (i.e. they are composed of identical $n \rightarrow k$ protocols). Note that Campbell et. al. in \cite{campbell} allows varying $k$ within a single factory, across different rounds.  


In performing our evaluations we consider four different system configurations: \emph{surplus} architectures that minimize application latency by setting the magic-state output capacity to the peak T gate request count in an application, \emph{singlet} architectures that minimize required space for the factory by producing only a single state per distillation cycle, \emph{optimized-unified} architectures that use one central factory with an optimized choice of output capacity $K$ and number of distillation rounds $\ell$, and \emph{optimized-distributed} architectures that choose an optimum output capacity $K$ distributed into an optimum number $X$ of factories, each utilizing $\ell$ distillation rounds. These architectures are summarized in Table \ref{tab:configurations}.

\begin{figure}[t!]
    \centering
    \includegraphics[width=\linewidth]{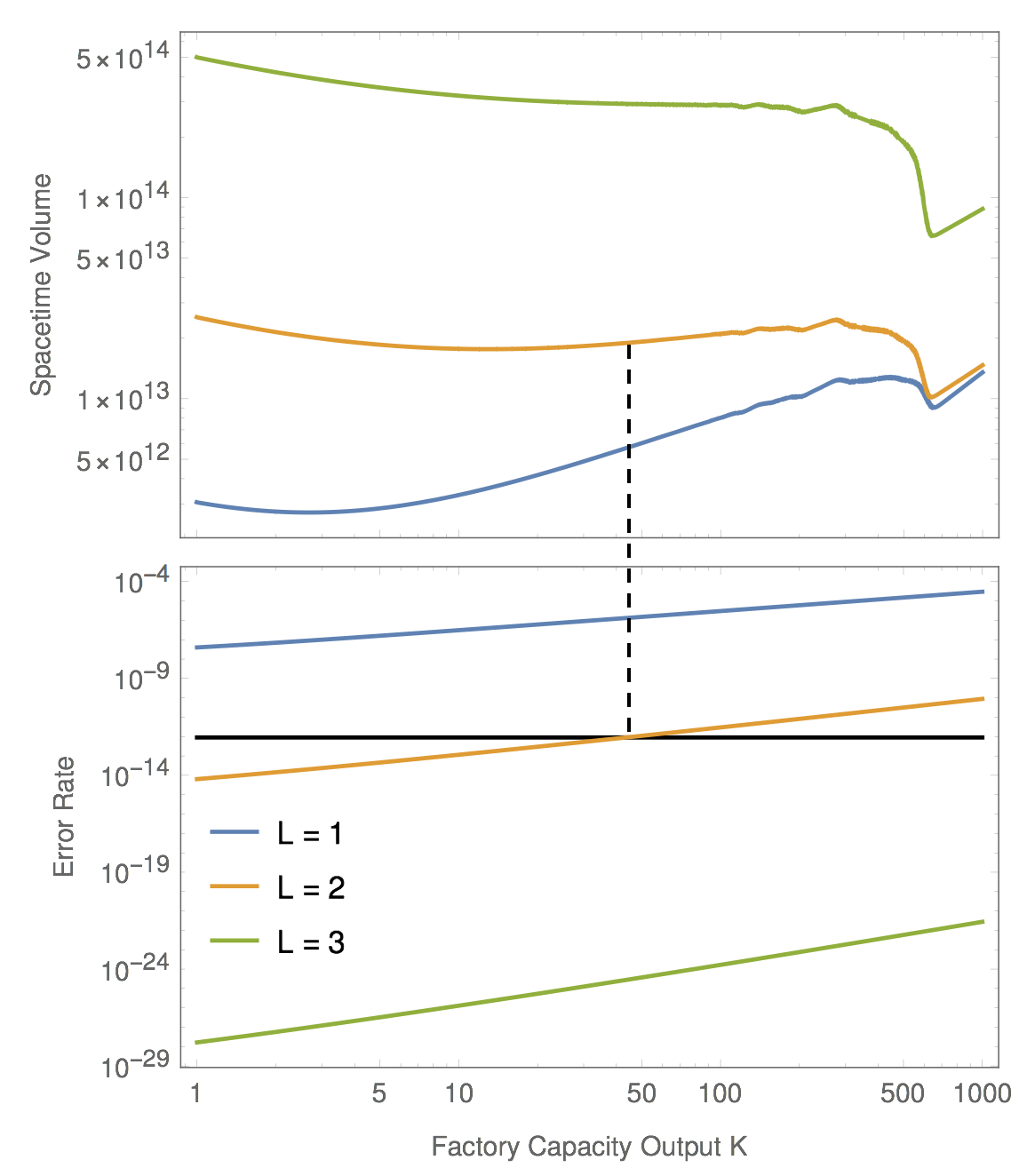}
    \caption{Space-time volume minimization under error threshold constraints imposed by target error rate for each block code level. An application will set a target error rate (black) that the factory must be able to achieve in output state fidelity. On the lower plot, levels 2 and 3 are the only levels available that can satisfy this. In the upper plot, we find that the lowest volume in the feasible area is located on the level 2 factory feasibility line. Recall the volume shapes are explained earlier in section \ref{sec:latency}. Here the tails after $K \approx 800$ show an increase in volume, as the added capacity grows the factory areas while maintaining constant latency.}
    \label{fig:errorfeasible}
\end{figure}

\subsection{Optimization Algorithm}\label{subsec:algo}
As keen readers may have already observed from Figure \ref{fig:fact_time} and Figure \ref{fig:area_law}, for fixed output capacity $K$, it costs us both in time latency and in factory footprint to implement a high $\ell$ block-code factory. The only reason we design for high $\ell$ is to achieve the desired target error rate. This relation is best captured in the bottom half of Figure \ref{fig:errorfeasible}, where the $L=1$ factory is not feasible for $K\ge 1$ since its output error rate is higher than the target error rate, while the $L=2$ factory is feasible for $K \in [1,50]$, and the $L=3$ factory is feasible for the entire plotted range.

We combine all of the details of the explicit overhead estimation derived above in order to find optimal design points in the system configuration space. To do this, we must ensure that designs are capable of producing the target logical error rate for an application. Additionally, there exists a set of constraints $C$ that $K,X \in \mathbb{Z^+}$ have to satisfy: (i) $1 \le X \le K$; (ii) $K/X \le (1-8\epsilon_{\text{inject}})/(3\epsilon_{\text{inject}})$, due the Bravyi-Haah protocol error thresholds. With the feasible space mapped out, standard nonlinear optimization techniques are employed to explore the space and select the space-time optimal design point. 
\begin{algorithm}[h]
		\caption{Space-time Optimization Procedure}
		\label{alg:opt}
		\algorithmicrequire{\;$P_s$, $N_{\text{gates}}$, $\epsilon_{\text{inject}}$, distribution $D$ and constraints $C$}
		
		\algorithmicensure{\; $K$, $X$}
		\begin{algorithmic}[1]
		\Procedure{Optimize}{}
		    \State $K \gets 1$, $X \gets 1$, $\ell_{\text{max}} = 5$
		    \State $\epsilon_{\text{target}} \gets P_s/N_{\text{gates}}$
			\For{$\ell \in [1,\ell_{\text{max}}]$}
			\State $k_\ell \gets (K/X)^{1/\ell}$
			\State $n_\ell \gets 3k_\ell + 8$
			\For{$r \in \{1, \cdots \ell\}$}
			    \If{$r==\ell$} $\epsilon_r \gets \epsilon_{\text{target}}$
			    \Else \;$\epsilon_r \gets (1+3k_\ell)^{2^r-1}\epsilon_{\text{inject}}^{2^r}$
			    \EndIf
			    \State $d_r \gets \text{Solve}\{d_r \cdot (100 \epsilon_{\text{in}})^{(d_r+1)/2} = \epsilon_r, \; d_r\}$
			\EndFor
			
				\State $R\equiv K/X \gets \text{Solve}\{\epsilon_{\ell} = \epsilon_{\text{target}},\;R\}$
				\If {$R \geq 1$}
				    \State $K_{\text{output}}\gets K \cdot \prod_{r=1}^{\ell}\big[(1-n_\ell\cdot\epsilon_{\text{inject}})\epsilon_r\big]$
				    \State $s \gets \lfloor t/K_{\text{output}}\rfloor$
				    \State $T_{\text{t}} \gets 4d_\ell + 4$
				    \State $T_{\text{distill}} \gets 11\sum_{r=1}^\ell d_r$
				    \State $n_{\text{distill}} \gets \frac{T_{\text{t}}}{T_{\text{distill}}} \cdot \Big(s \cdot \sqrt{\frac{K}{X}} +  \sqrt{\frac{t-sK}{X}}\Big)$
				    \State $T_{\text{total}} \gets T_{\text{distill}}\Big(\sum_{t=0}^{T_{\text{peak}}}n_{\text{distill}}\Big)\cdot D[t]$
				    \State $A_{\text{factories}} \gets X\cdot n_l^{\ell-1}\cdot(6k_\ell + 14)\cdot d_1^2$ 
				    \State $(K,X) \gets \;\argmin_{(K,X):C}\; A_{\text{factories}} \cdot T_{\text{total}}$
				\Else
				    \State $\ell \gets \ell + 1$
				\EndIf
			\EndFor
			\State \Return $K,X$
		\EndProcedure
		\end{algorithmic}
	\end{algorithm}
	
With these constraints in mind, we explore the space by first selecting the lowest $\ell$ possible. As the area law and full volume scaling trends of the previous sections indicate, if there are any feasible design points with $\ell = \ell_0$, then any feasible design points for systems with $\ell_i > \ell_0$ will be strictly greater in overall volume. This is somewhat intuitive, as concatenation of block code protocols is very costly.

With the lowest $\ell$ selected, we check to see if there exists any feasible design points for this $\ell$ by checking for solutions to the equation:
\begin{align}
    (1+3k^{\sfrac{1}{\ell}})^{2^{\ell}-1}\epsilon_{\text{inject}}^{2^{\ell}} \leq \frac{P_s}{N_{\text{gates}}}
\end{align}
If the $K$ that solves this equation is greater than or equal to 1, then there does exist feasible design space along this $\ell$, and the algorithm continues. Otherwise, $\ell$ is incremented. 

Next, nonlinear optimization techniques are used to search within the mapped feasible space for optimal design points in both $K$ and $X$.

\subsection{Simulation and Validation}\label{sec:simulation}
\begin{figure}[t]
  \centering  
  \includegraphics[trim={0cm 0cm 0cm 0cm}, width=\columnwidth]{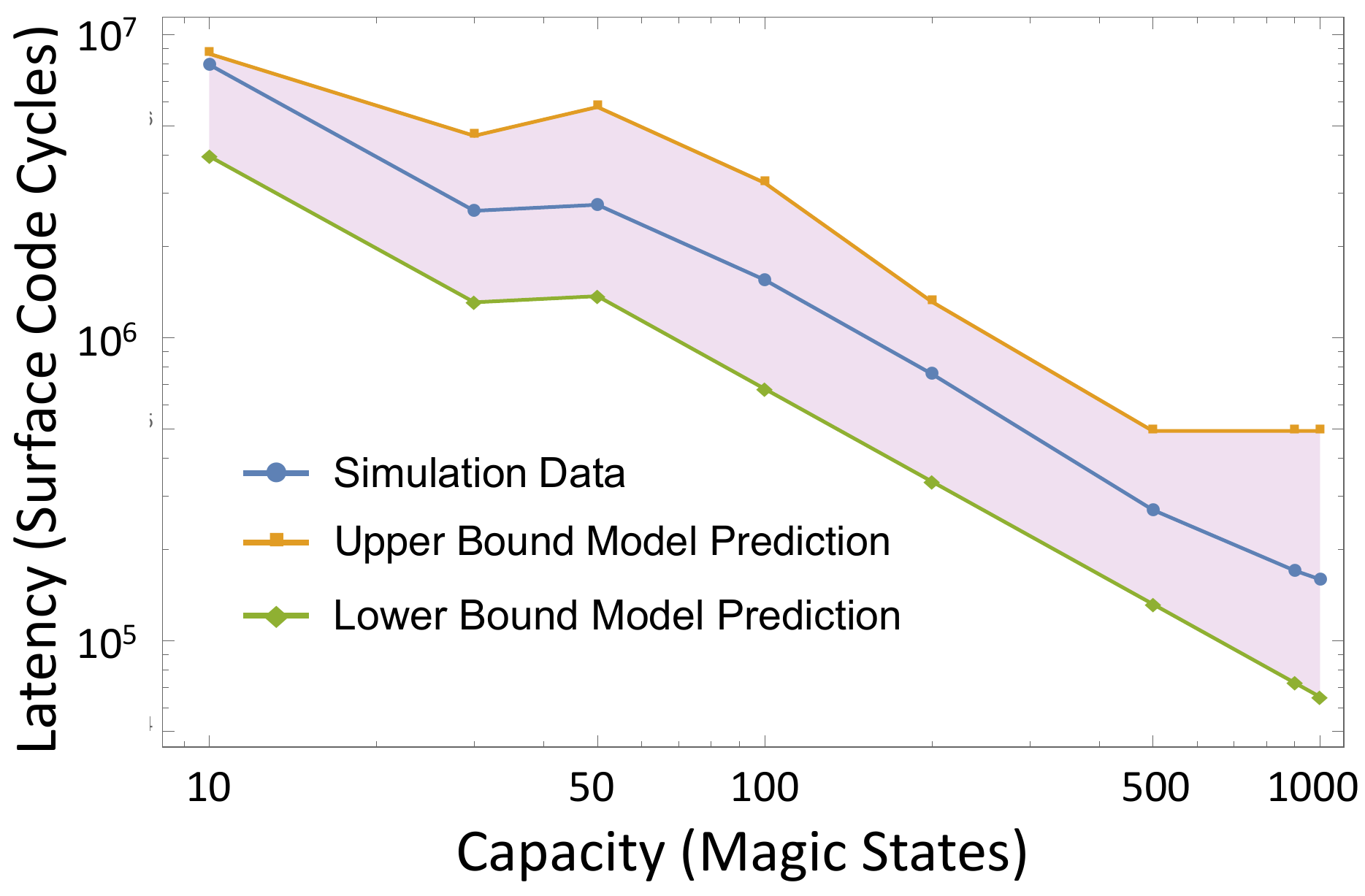}
  \caption{Model validation by simulation. The simulation data (blue line) lies between the upper bound model prediction that overestimates congestion(orange line), and the congestion-free lower bound (green line). }
  \label{fig:simwithlower}
\end{figure}

This section explores the validity of our models through empirical evaluation of the space-time resources. To do this, we improve the surface code simulation tool from~\cite{javadi2017optimized} to accurately assess the latency and qubit cost of fully error-corrected applications with various magic-state distillation factory configurations. Specifically, we added support for arbitrary factory layouts, which manifests as black boxed regions dedicated to factories that cannot be routed through during computation, combined with sets of locations of produced magic states. The result is a cycle precise simulator that accurately performs production and consumption of magic states, including all necessary routing. 

One implementation detail that is supported is the ability to dynamically reallocate specific magic-state assignments during runtime. Statically, each T gate operation is prespecified with a particular magic-state resource, located along the outer edge of a factory. During runtime, this can introduce unnecessary contention, as two nearby logical qubits can potentially request the same magic state. This is avoided by implementing online magic-state resource shuffling, so that if the particular state that was requested is unavailable, the system selects the next nearest state that is available. If no such states exist, this T gate is stalled until the next distillation cycle is completed.

Figure~\ref{fig:simwithlower} shows simulation results superimposed on top of those driven analytically. We can see that the model shows the same trend as the simulation behavior (blue line), and thus we will be able to show relative tradeoffs between capacity and latency. For simplicity the validation is performed on single unified factory located at the center of the surface code lattice. The results extend well to multiple factories, because in the distributed case, each factory will be responsible for magic-state requests in a sub-region of the lattice.

\begin{figure}[t]
  \centering  
  \includegraphics[trim={0cm 0cm 0cm 0cm}, width=\columnwidth]{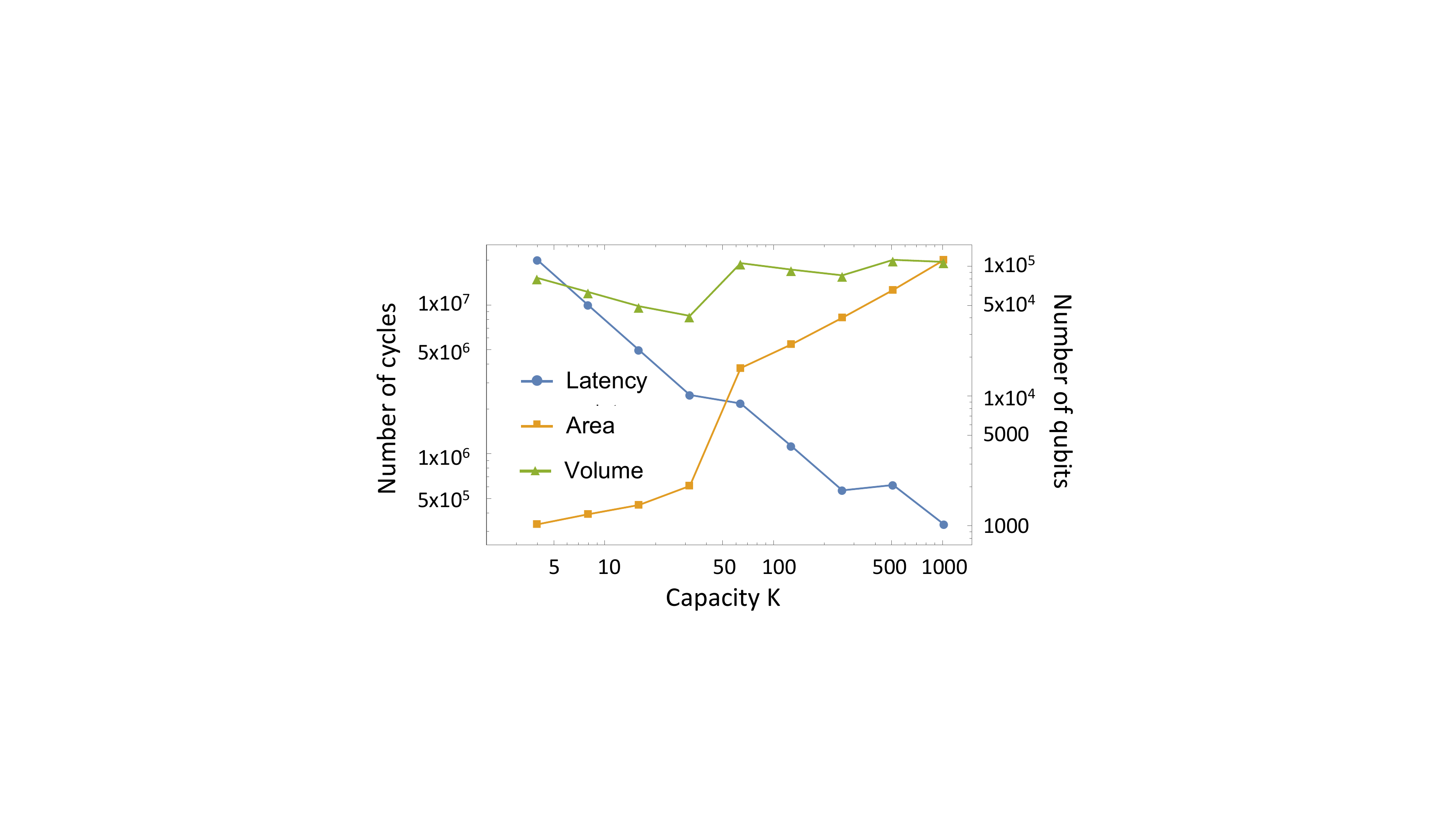}
  \caption{Space-time tradeoff observed empirically in simulation for varying factory capacities. A space-time volume (green line) can be chosen at $K\approx40$, which is an optimal, minimized value on this curve. It corresponds here to a low-qubit, high latency configuration. Notice that another configuration at $K\approx300$) could be chosen, corresponding to a high-qubit, low-latency configuration. In this case, the former of these choices is more resource optimized, as the space-time cost is lower.}
  \label{fig:xmen}
\end{figure}

We can validate this by simulating optimal operating points in the space-time trade-off spectrum and comparing them to our expectation from the model. Using simulation data, we re-plot our idealized tradeoff in Figure~\ref{fig:tradeoffcartoon} for the Ising Model Application and show the results in Figure~\ref{fig:xmen}. We see that as factory capacities increase, the applications time improves at the expense of its qubit numbers. In this figure, the space-time volume is sketched in green, and has two near-optimal points: one with relatively few qubits but high latency, and vice versa. The worst performance occurs in the middle of this spectrum, when transition from level 1 to level 2 distillation needs to occur (causing a sudden jump in qubits, but not much latency improvement).

\section{Results}\label{sec:results}
In this section we present the resource requirements of various magic-state factory architectures, and show that by considering the scaling behaviors that we have highlighted and searching the design space with our optimization algorithm, we can discover system configurations that save orders of magnitude of quantum volume. 

We first compare the overheads of the surplus and singlet architectures that represent baselines against which we compare our optimized architectures. We then compare the surplus architecture with the optimized-distributed design found with our optimization algorithm. We look at two representative benchmarks for the quantum chemistry and quantum simulation fields, the Ising Model \cite{barends2016digitized} and Ground State Estimation \cite{whitfield2011simulation} algorithms, as well as how performance of these architectures changes as the benchmarks scale up in size. Next, we detail the space and time trade-off that is made in our resource optimized design choices, and show that the latency induced by a design is a more dominant factor in these applications. We then present a full design space comparison, showing the performance of the surplus design against the singlet design, as well as the optimized-unified factory design, all compared to the performance of optimized-distributed design. Lastly, we analyze the sensitivity of these designs to fluctuations in the underlying physical error rates, and show that building out a distributed factory design adds robustness that makes the architecture perform well for a wider range of input parameters.

\subsection{Comparing Surplus and Singlet Architectures}
\begin{figure}[t]
    \centering
    \includegraphics[width=\linewidth]{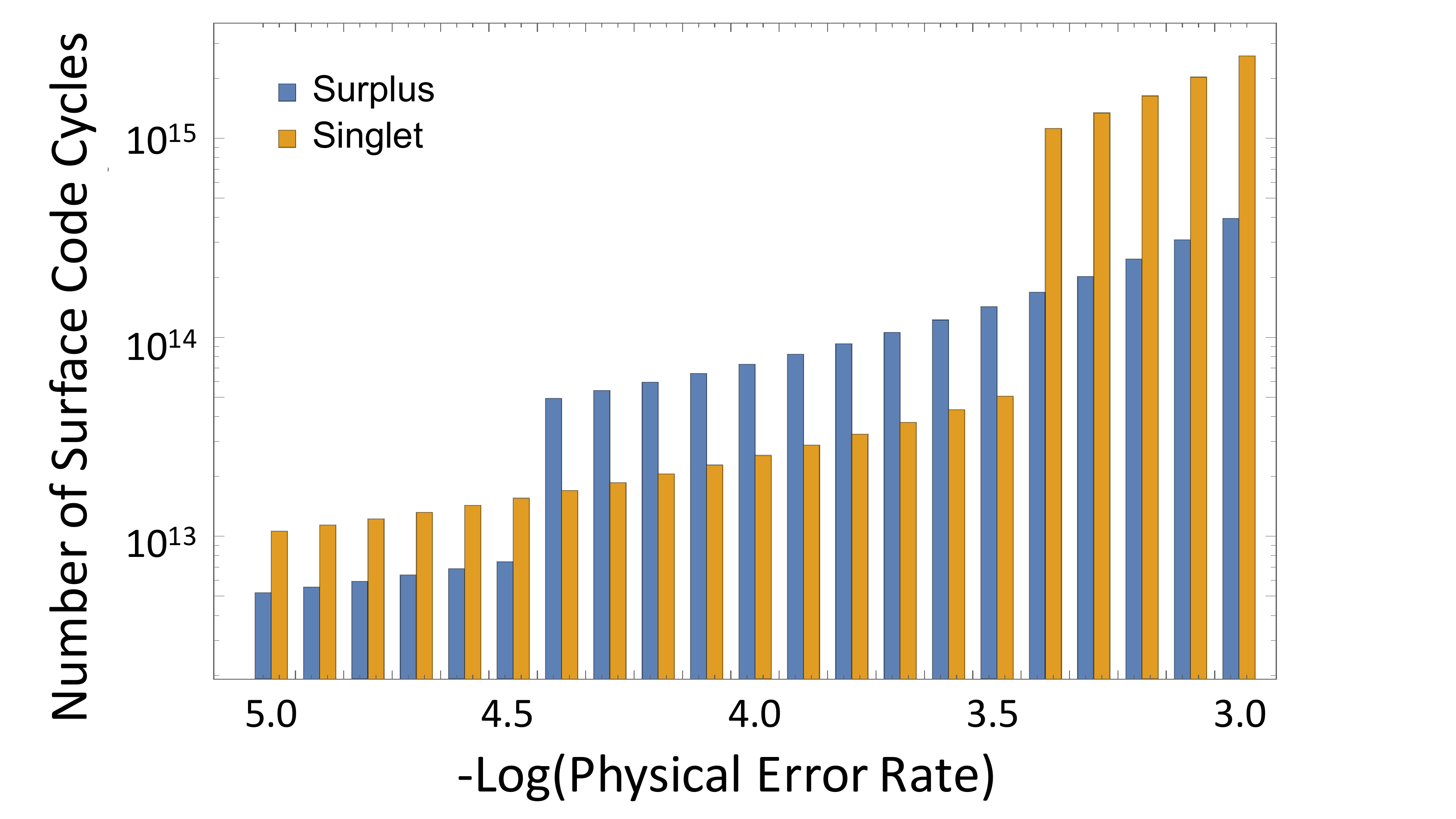}
    \caption{(Color online) Comparing surplus and singlet designs. There are regions where each outperforms the other, showing great sensitivity to the underlying physical error rate and the corresponding required $\ell$. Recall that the step-like shape is due to level transitions explained in section \ref{subsec:algo}.}
    \label{fig:timespaceoptimal}
\end{figure}

We begin with Figure \ref{fig:timespaceoptimal} by comparing two architectures that aim solely to minimize application latency or required space. This comparison represents the range between two ends of the design space spectrum for single factory architectures, and each shows a particular error rate range over which it performs more optimally. Initially, at the highest input error rate, the space optimal singlet design requires more resources than the time optimal surplus design, as the application suffers from excessive latency from magic-state factory access time. Note the inflection points at $10^{-3.5}$ and $10^{-4.5}$ input error rates. At these points, the singlet factory is able to reduce the number of rounds of distillation it must perform, as input error rates are sufficiently low. Over this region, the reduction in area compensates the expansion in computation time, and the design outperforms the much larger surplus factory configuration. At $10^{-4.5}$, the surplus factory is able to operate with fewer distillation rounds as well, enabling this configuration to outperform the singlet design.

This behavior is surprising, as it indicates that with respect to a high-parallelism application, there are input error rate regions where intuitively conservative, space minimizing designs are able to outperform what seem like aggressively optimized designs. We see this because we are comparing space and time simultaneously, which allows us to see that the trade-off is asymmetric and these factors interact non-trivially. 

\subsection{Optimized Design Performance}

\begin{figure*}[t]
    \centering
    \begin{subfigure}[b]{0.47\textwidth}
    \includegraphics[width=\linewidth]{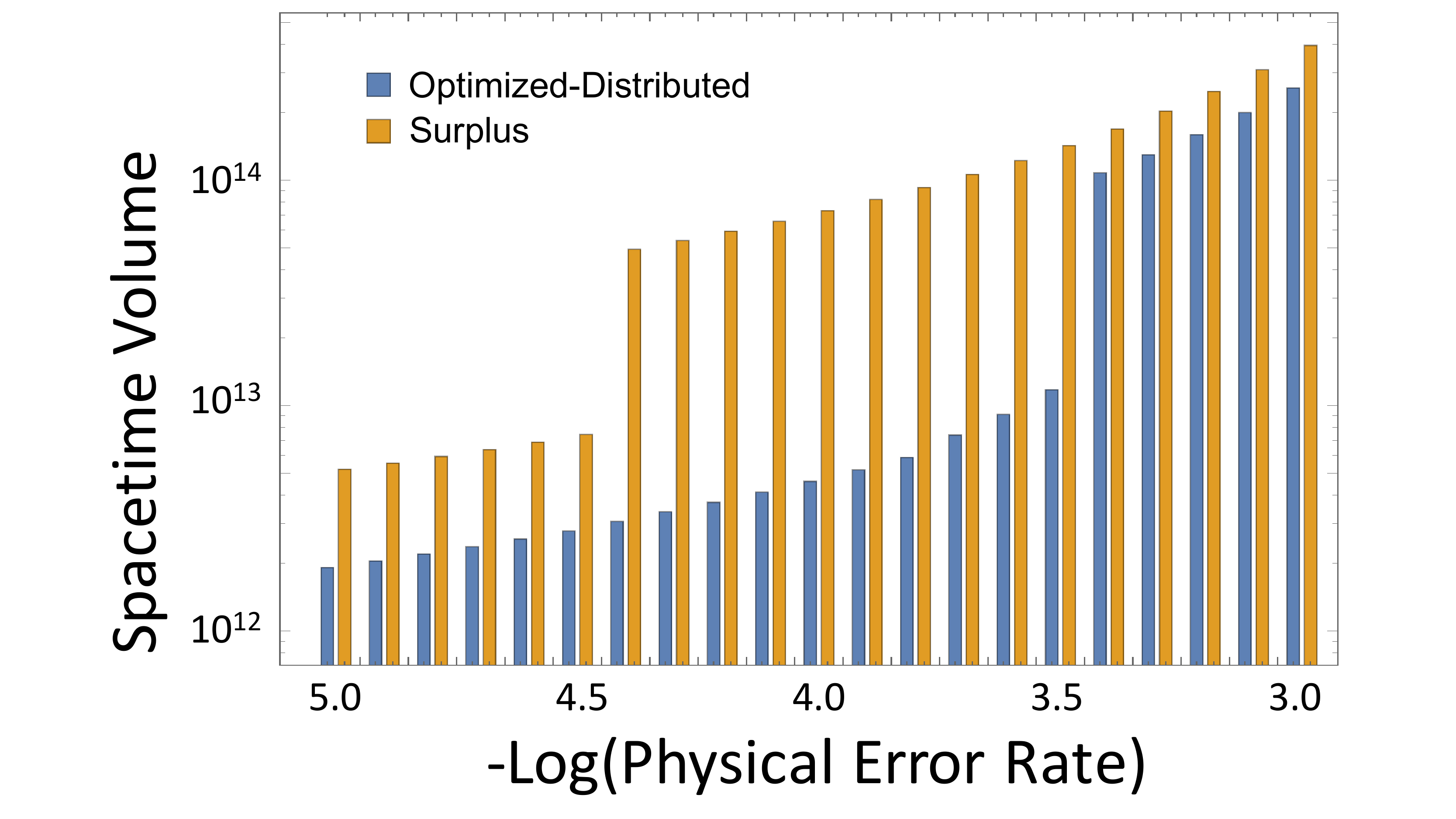}
    \caption{Ising Model N=500}
    \label{fig:isingresults}
    \end{subfigure}
    \begin{subfigure}[b]{0.47\textwidth}
    \centering
    \includegraphics[width=\linewidth]{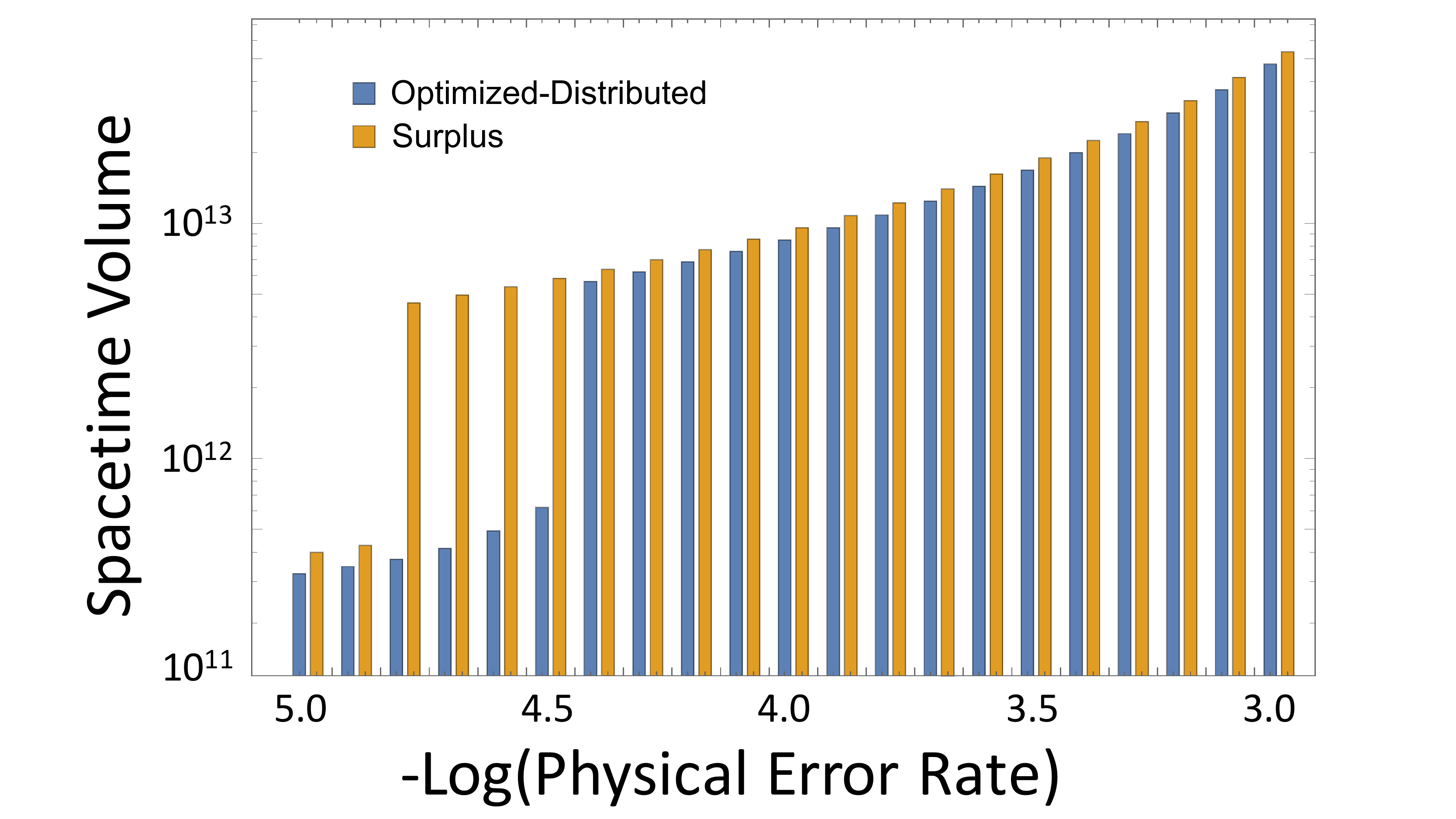}
    \caption{Ground State Estimation}
    \label{fig:gseresults}
    \end{subfigure}
    \begin{subfigure}[b]{0.47\textwidth}
    \includegraphics[trim={0 0 0 1cm}, width=\linewidth]{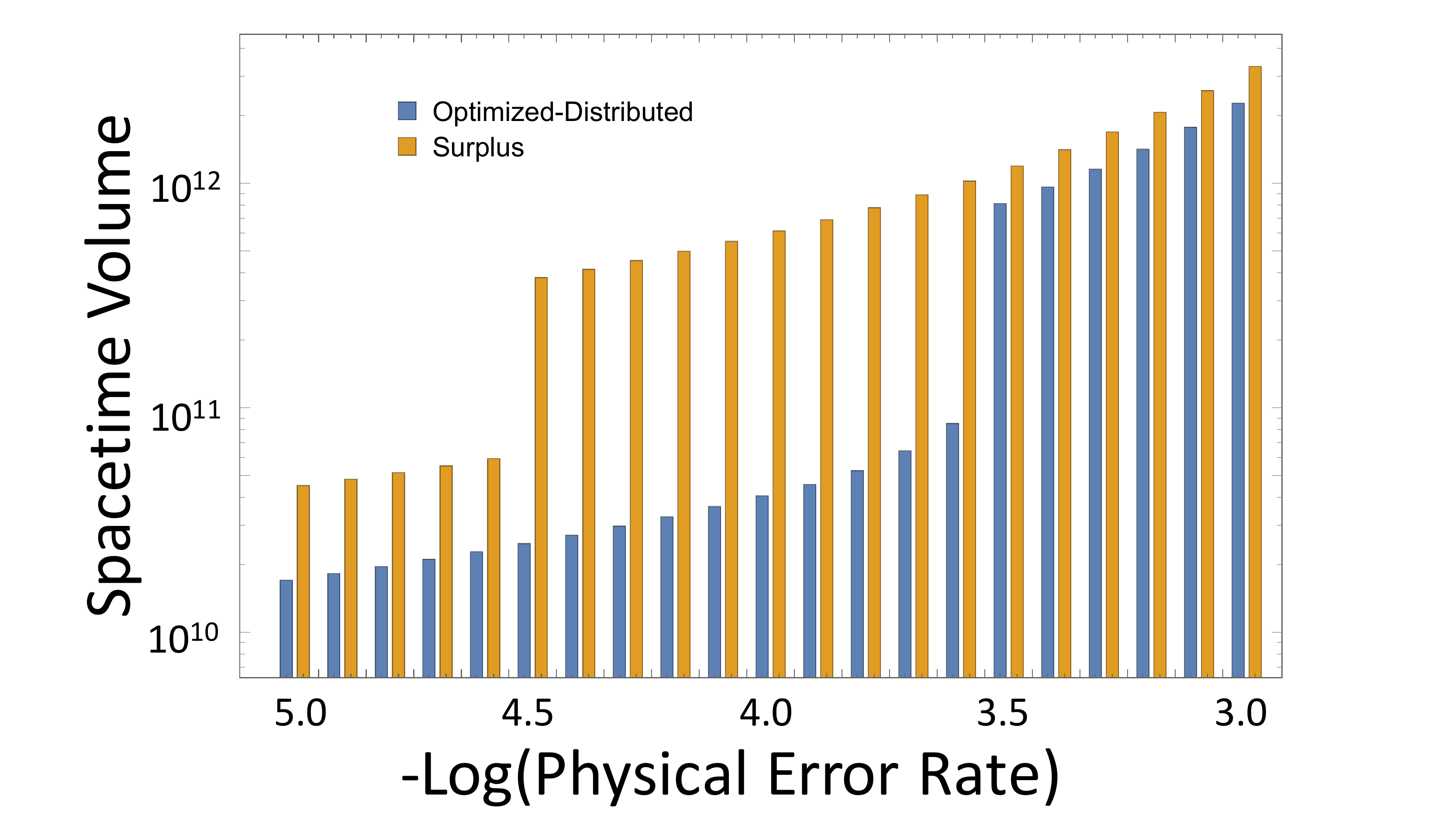}
    \caption{Ising Model N=1000}
    \label{fig:ising1000}
    \end{subfigure}
    \begin{subfigure}[b]{0.47\textwidth}
    \centering
    \includegraphics[width=\linewidth]{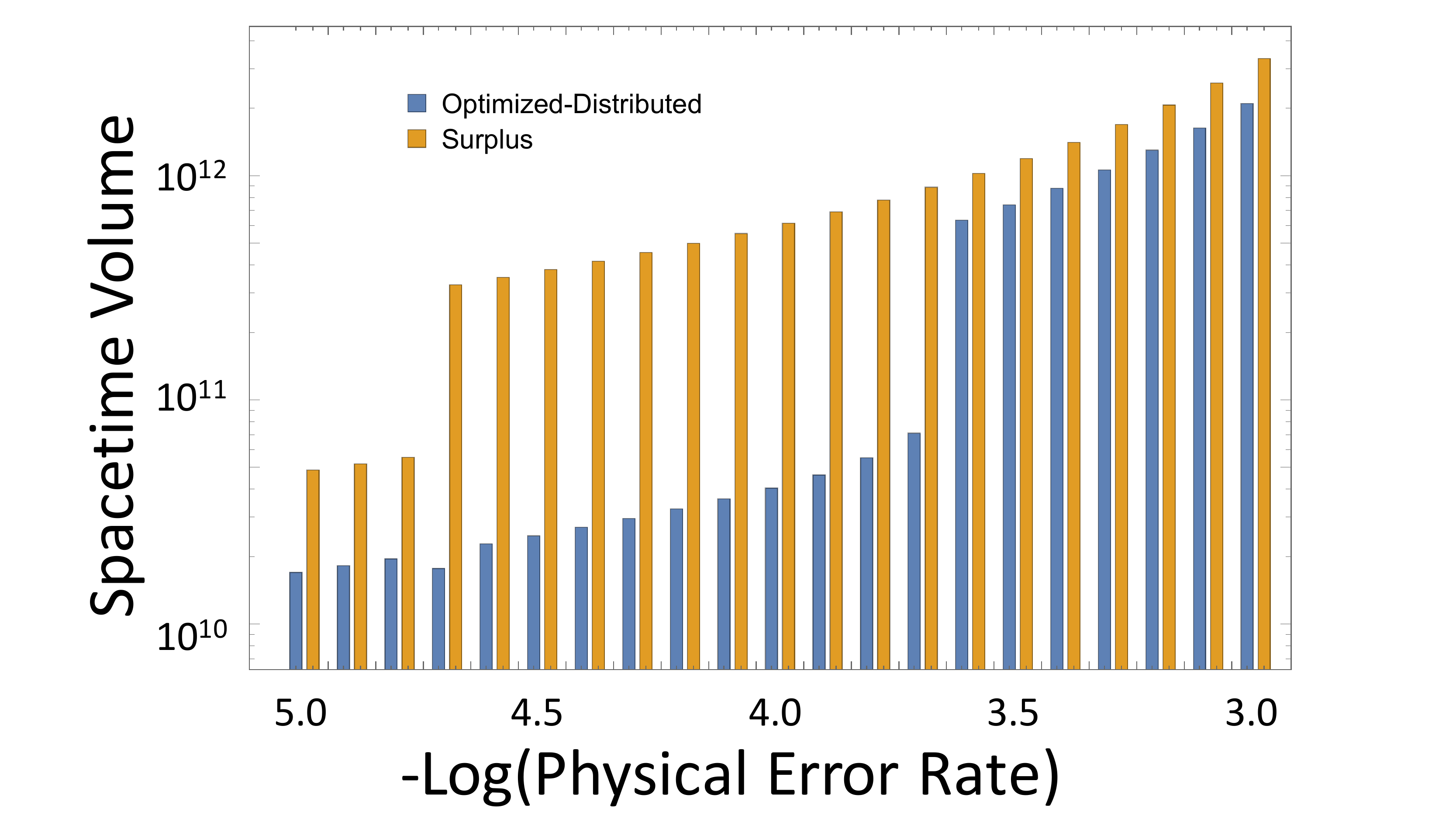}
    \caption{Ising Model N=2000}
    \label{fig:ising2000}
    \end{subfigure} 
    \caption{(Color online) (a)-(b) Resource reductions of optimized-distributed  designs over surplus designs for both Ising Model and Ground State Estimation. While Ising Model is intrinsically more parallel which leads to high choices of output capacity, both applications still show between a 12x and 16x reduction in overall space-time volume. (c)-(d) Ising Model with varying problem sizes, comparing time optimal factories against fully space-time optimized configurations. We see that the trend of between 15x and 20x total volume reduction extends to larger molecular simulations.}
    \label{fig:appresults}
\end{figure*}

\begin{figure*}[t!]
    \centering
    \begin{subfigure}[b]{0.33\textwidth}
    \includegraphics[width=\linewidth]{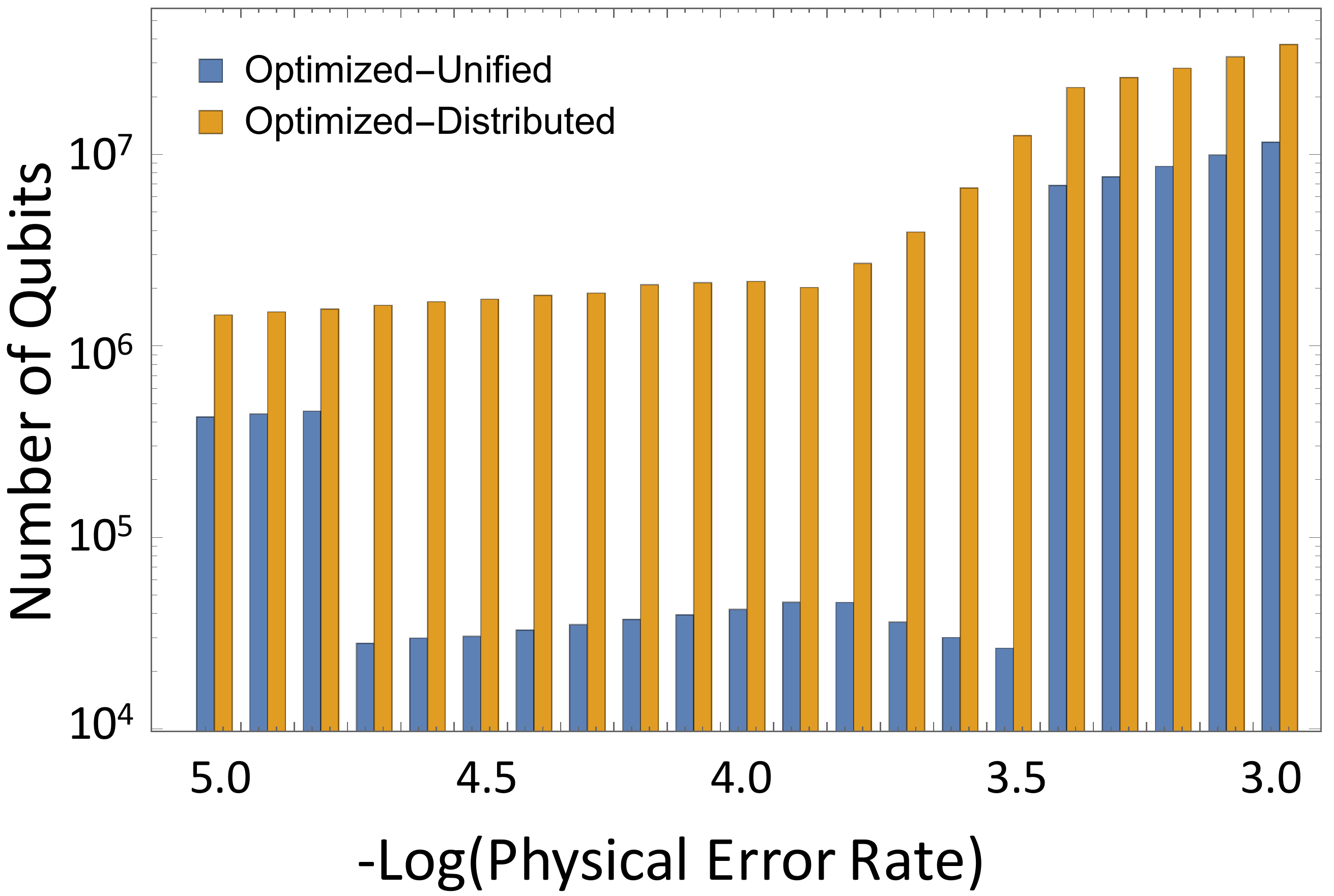}
    \caption{Space tradeoff}
    \label{fig:spacereduction}
    \end{subfigure}
    \begin{subfigure}[b]{0.33\textwidth}
    \centering
    \includegraphics[width=\linewidth]{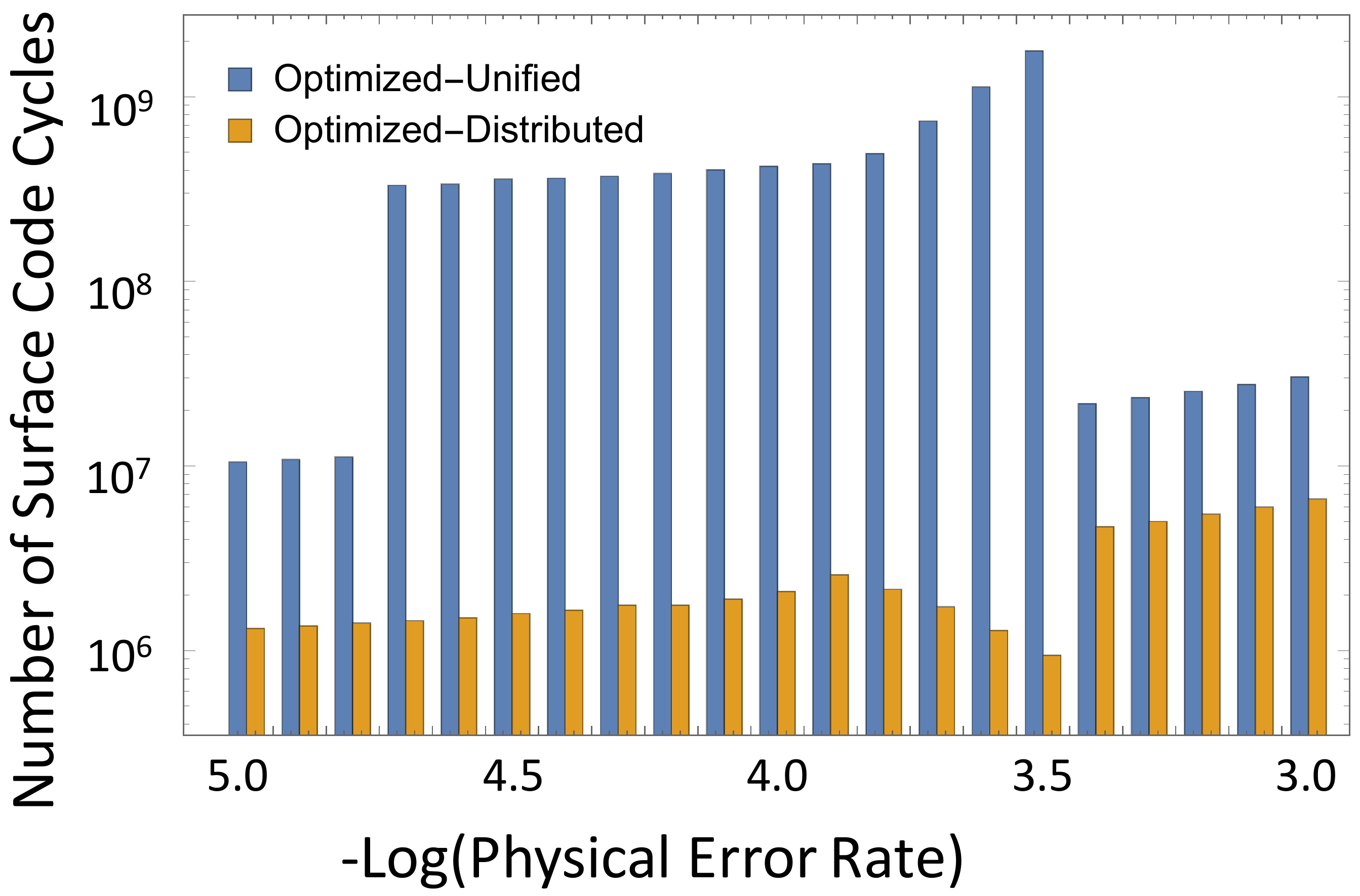}
    \caption{Time tradeoff}
    \label{fig:timereduction}
    \end{subfigure}
    \begin{subfigure}[b]{0.33\textwidth}
        \includegraphics[width=\linewidth]{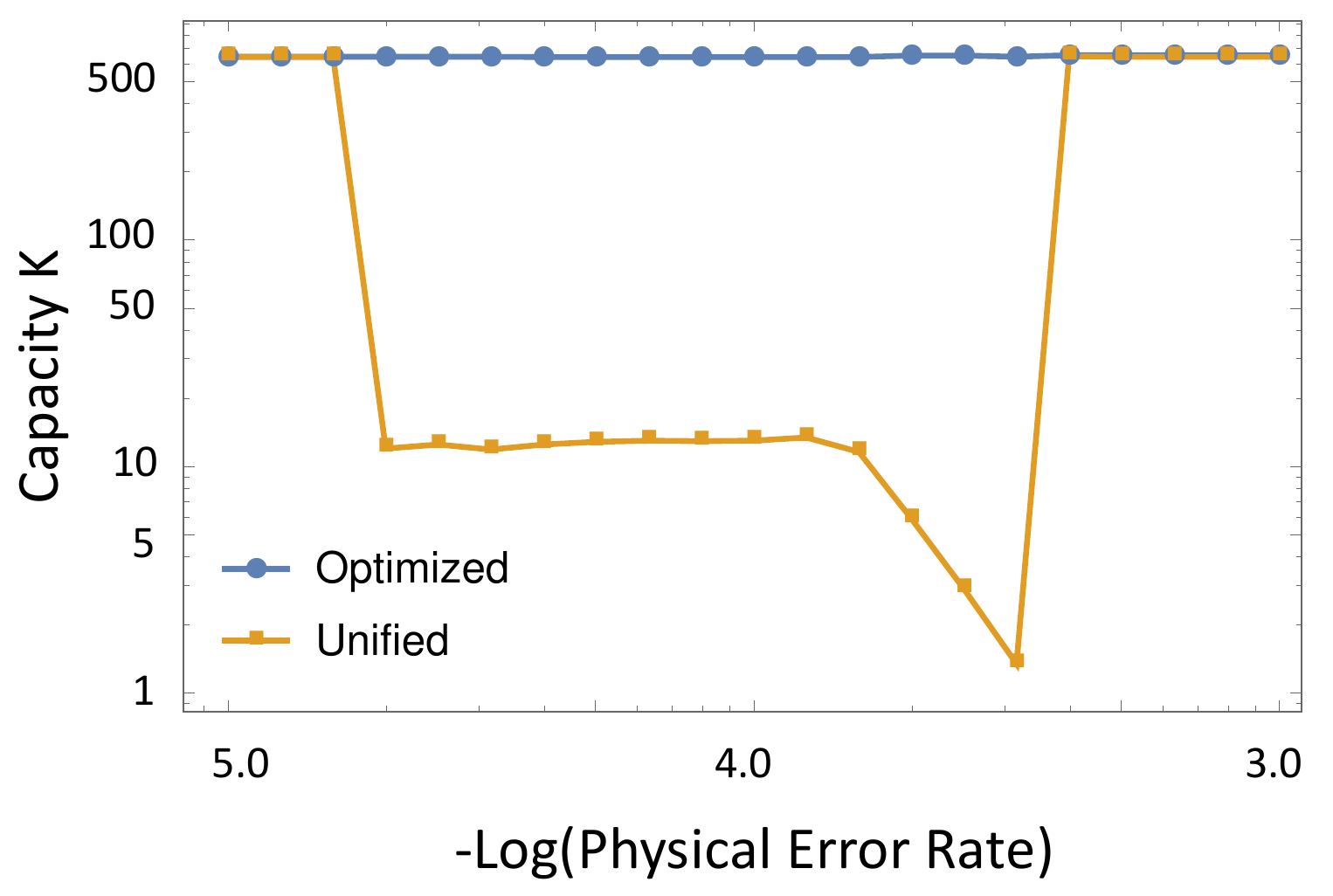}
    \caption{Output capacities procedurally selected}
    \label{fig:kvals}
    \end{subfigure}
    \caption{(Color online) Space-time volume reduces by moving from an optimized-unified factory to an optimized-distributed factory, as the designs trade space for time. Magic-state access latency is a dominating effect in these applications, as can be seen by the large capacity values chosen by the optimized factory configuration.}
    \label{fig:optimizedreductionresults}
\end{figure*}

We now move to comparing the surplus design against the optimized-distributed design discovered by our optimization algorithm, that is allowed to subdivide factories across the machine. Figures \ref{fig:isingresults} and \ref{fig:gseresults} depict the detailed results of our optimization procedure on the Ising Model and Ground State Estimation applications, respectively. Ising Model is intrinsically very parallel, which leads to a higher optimal capacity choice for the optimized-distributed factory. Note however that it is able to choose a distribution level that saves approximately 15x in space-time volume. Ground State Estimation is very serial, yet for sufficiently low error rates the optimized-distributed design is able to incorporate distribution of factories into the lattice to lower the required block code concatenation level $\ell$, resulting in a 12x reduction in volume across these points.

The reason that the distributed factory design is able to outperform the surplus design is that the feasibility regions of the two designs differ. Because the distributed factory utilizes many small factories on the machine it can achieve a higher output state fidelity than a single factory design, which enables it to operate with a smaller number of distillation rounds. The optimization algorithm respects this characteristic, which is why it searches iteratively from the lowest number of distillation rounds possible, one by one until it discovers a feasible factory configuration.

\subsubsection{Optimized Design Performance Scaling}
Figures \ref{fig:ising1000} and \ref{fig:ising2000} detail these trends as larger and larger quantum simulation applications are executed. For extremely large simulations, we find that the volume reductions that optimizing a factory design yields become even more pronounced, resulting in between a 15x and 18x full resource reduction. These designs also show sensitivity to physical error rates that require designs to change block code distillation level.

\subsection{Distributed Factory Characteristics}
As Figure \ref{fig:spacereduction} describes, an optimized-distributed set of factories is able to save between 1.2x and 4x in total space-time volume over the optimized-unified factory. Large volume jumps occur primarily between $10^{-3.5}$ and $10^{-3.4}$ physical error rate, and this again corresponds to a requirement by this application to increment to a higher block code level $\ell$, which happens for both the unified and distributed factory schemes. 

These optimized designs trade space for time, as Figures \ref{fig:spacereduction} and \ref{fig:timereduction} indicate, and the net effect is an overall volume reduction. This is indicative that for these highly parallel quantum chemistry applications, the magic-state factory access latency is a much more dominating effect than the number of physical qubits required to run these factories. 

Figure \ref{fig:kvals} depicts the output capacities chosen by the optimization procedure, and how they differ when the system is unified or distributed. Notably, at both ends of the input error rate spectrum we find that both factory architectures choose the same output capacity, as in the high error rate case this is driven by high $\ell$ requirement, while in the low error rate limit both factory architectures can afford to be very large and not suffer from any yield penalties. However, through the center of the error rate spectrum the unified factory design must lower the chosen output capacity, as supporting higher capacity would require a very expensive increase in the number of distillation rounds. 

\subsection{Full Design Space Comparison}
\begin{figure*}[t!]
    \centering
    \includegraphics[width=0.8\linewidth]{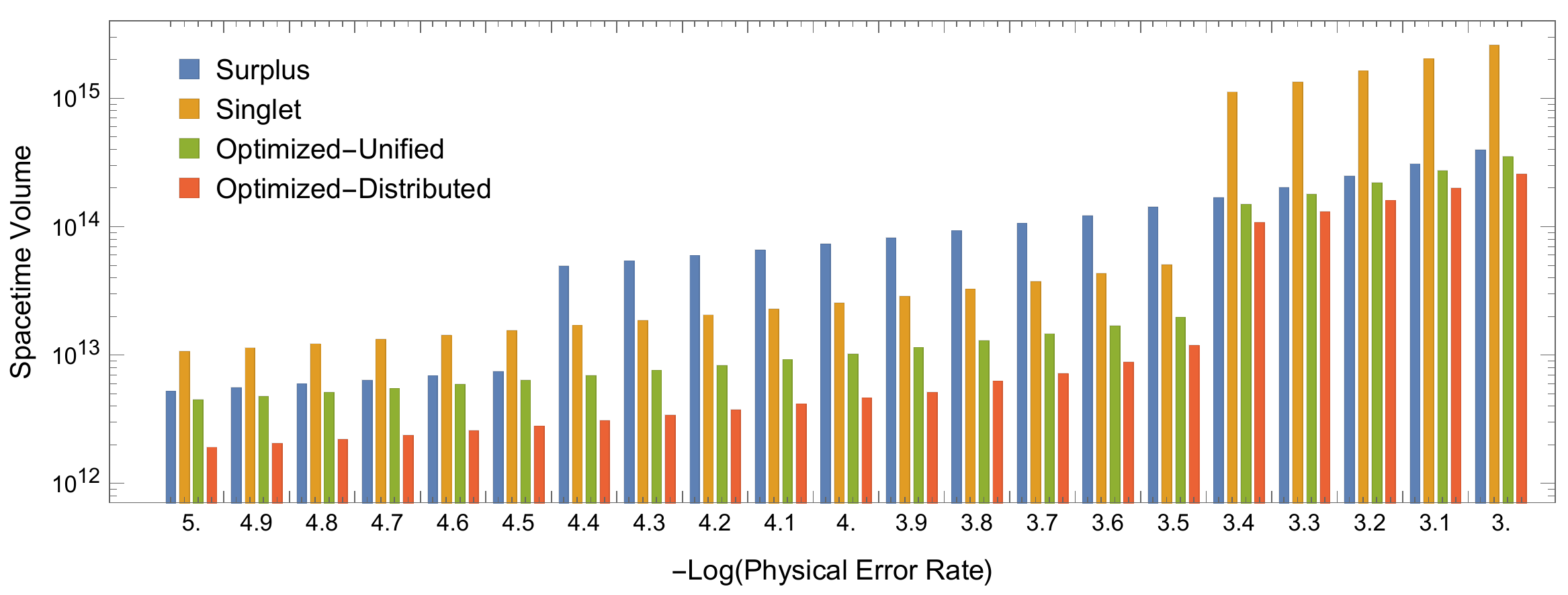}
    \caption{(Color online) Full volume comparison across distillation factory architectures.} 
    \label{fig:fullvols}
\end{figure*}

Figure \ref{fig:fullvols} depicts the full space-time volume required by different factory architectures across the design space. Shown are the four main configurations: a surplus factory configured with output capacity $K = T_{\text{peak}}$, a singlet factory with $K=1$, an optimized-unified factory, and an optimized-distributed factory.

Distinct volume phases are evident visually on the graph, due to the different feasibility regions of the architectures. Sweeping from high error rates to low error rates, large volume jumps occur as observed before, for specific configurations when that configuration can operate with fewer rounds of distillation in order to convert the input error rate to the target output error rate. Notice that this jump occurs earliest for the singlet, optimized-unified, and optimized-distributed designs, at $10^{-3.5}$ input error rate. All of these designs show an inflection point here, where the configurations can achieve the target output error rate with a smaller number of block code distillation levels. This is not true of the surplus factory, which in fact has the largest output capacity of the set. Because the output capacity is so high, the lowest achievable output error rate is much higher than that of the other designs. This forces the block code level to remain high until the input error rate becomes sufficiently low, which occurs at $10^{-4.5}$. 
~\\\subsection{Sensitivity Analysis} \label{sec:sensitivity}
\begin{figure}[t!]
    \centering
    \includegraphics[width=\linewidth]{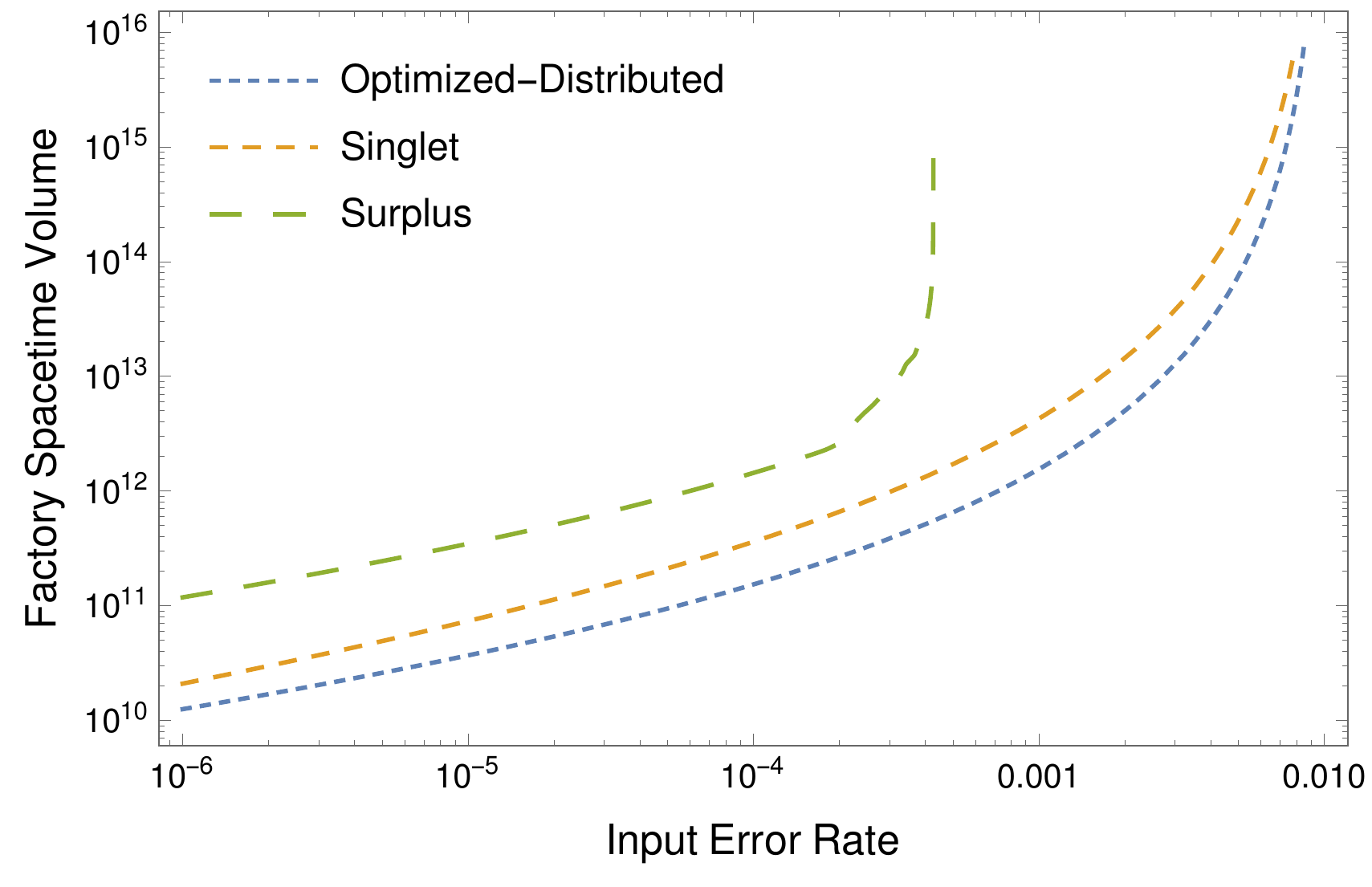}
    \caption{Factory architectures and their sensitivities to fluctuations in underlying physical error rates}
    \label{fig:sens}
\end{figure}

Now we turn to analyzing how these designs perform if the environment in which they were designed changes. Supposing that a design choice has been made specifying the desired factory capacity $K$, number of factories $X$, and block code distillation level $\ell$, different types of architectures show varying sensitivity to fluctuations in the underlying design points around which the architectures were constructed. For example, Figure \ref{fig:sens} details an instance of this occurrence. The figure shows the surplus, singlet, and optimized-distributed factory designs, in this case setting $K\sim 600$ and $X \sim 200$ for the distributed architecture. All of these factories were designed under the assumption that the physical machine will operate with $10^{-5}$ error rate.

We see that while these applications perform similarly over the range from $10^{-5}$ to $10^{-4}$, just after this point the surplus factory encounters a steep volume expansion due to the yield threshold equation \ref{eq:yieldthresh}. For this design the threshold of tolerable physical error rates is quite high, significantly higher than that of the other designs. Because of this, it can tolerate a smaller range of fluctuation in the underlying error rate before it ceases to execute algorithms correctly.

\section{Conclusion}\label{sec:conclusion}
We present methods for designing magic-state distillation factory architectures that are optimized to execute applications that present with a specific parallelism distribution. By considering applications with different levels of parallelism, we design architectures to take advantage of these characteristics and execute the application with minimal space and execution time overhead.

By carefully analyzing the interaction between various magic-state factory characteristics, we find that choosing the most resource optimized magic-state distribution architecture is a complex procedure. We derive and present these trade offs, and compare the architectures that have been commonly described in literature. These comparisons show a surprising picture: namely that even a modest factory capable of producing just a single resource state per distillation cycle can outperform the more commonly described surplus factory in particular input error rate regimes. We also propose a method of distributing the total number of magic states to be produced into several smaller factories uniformly distributed on a machine. In doing this, we see that these types of architectures are capable of achieving higher output fidelities of their produced states with added resilience against fluctuations of the underlying error rate, when compared to unified architectures composed of a single factory. While these designs are tailored to specific applications, we conjecture that distributed systems would in fact be more flexible in their abilities to execute applications with different amounts of parallelism. Intrinsic to their design is the ability to optionally compile smaller applications to various subunits of the machine. Because of this, these designs can be used to support a much wider range of application types than those comprised of a single factory.

These systems also show that the trade off in space and time is asymmetric. In quantum chemistry and simulation applications, we notice that the resource optimized designs can use upwards of 2 orders of magnitude more physical qubits to be implemented, while they end up saving over 3 orders of magnitude in time. Magic-state access time, or latency induced specifically by delays due to stalling as magic states are produced, we find is a dominating effect in the execution of these applications. In order to mitigate these effects in a resource-aware fashion, designing a distributed system of several factories allows for efficient partitioning of the magic-state demand across the machine, at the cost of physical area. 

These conclusions can have physical impacts on near-term designs as well. Specifically, the construction of a factory architecture can imply the location of physical control signals on an underlying device. What we are showing then is the effect of several theoretical long-term designs, and the conclusion that distributed sets of factories outperform other designs should help motivate device fabrication teams as they decide which physical locations should be occupied by rotation generating control signals. As a general principle, long term architectural design and analysis can help guide the study and development of near term devices, which ultimately will help hasten the onset of the fault-tolerant era \cite{preskill2018quantum}.
\section{Future Work}\label{sec:future}
There are a number of immediate extensions to this study: 
\begin{itemize}
\item \emph{Comparing distributed factory topologies.} Choosing an optimal layout for a distributed factory design is potentially very difficult, and requires an ability to estimate the overheads associated with different layouts. Using architectural simulation tools and adapted network simulation mechanisms, we can foresee evaluation of two new architectures: peripheral and asymmetric-mesh placement. Peripheral placement refers to factories surrounding a central computational region, while asymmetric-mesh placement refers to embedding the factories throughout the machine itself. 
\item \emph{Embedding data qubits within magic-state factories.} While the designs presented here assume that magic-state factory regions are to be considered black boxes that are not to be occupied by data qubits, because of their massive size requirements we imagine a system that embeds the relatively smaller number of data qubits within the factories themselves. A study of the effect of various embedding techniques on factory cycle latency could determine the efficiency of such a design. 
\item \emph{Advanced factory pipeline hierarchy.} We envision a concatenation of clusters of the magic-state factories, targeting continuous outputs in time, and hence reduction in contention caused by the distillation latency. In particular, each sub-region in the mesh contains multiple small, identical factories that were turned on asynchronously. So at each time step, there will always be a factory that completes a distillation cycle, and thus serving magic state continuously.
\item \emph{Generalization to other distillation protocols.} Although the Bravyi-Haah protocol studied in this paper is among the best known protocols, little analysis has been done on other techniques discovered recently \cite{haah2017magic}. 
\item \emph{Optimizing the internal mapping and scheduling of magic-state factories.} This work has modeled factories as black-boxed regions that continuously produce resources. A realistic implementation of those factories that optimize for internal congestion would significantly reduce factory overhead, in conjunction with designs proposed in this work that optimize for external congestion. This was studied in \cite{ding2018magic}.
\item \emph{Flexibility of Distributed Magic-State Architectures.} While these designs are tailored to applications of a certain parallelism distribution, a study could analyze designs that balance domain specific optimization against general application compatibility. 
\end{itemize}

\section*{Acknowledgements}
This work was funded in part by NSF Expeditions in Computing grant 1730449, Los Alamos National Laboratory and the U.S. Department of Defense under subcontract 431682, by NSF PHY grant 1660686, and by a research gift from Intel Corporation.

\bibliographystyle{ieeetr}
\bibliography{references}

\end{document}